\documentclass{article}
\usepackage[margin=1in]{geometry}
\usepackage[utf8]{inputenc}
\usepackage[T1]{fontenc}    %
\usepackage{hyperref}       %
\usepackage[numbers,super,comma,sort&compress]{natbib}

\usepackage{authblk}
\usepackage{microtype}      %
\usepackage{lmodern}

\usepackage{color}
\usepackage{url}            %
\usepackage{booktabs}       %
\usepackage{amsfonts}       %
\usepackage{amsmath}        %
\usepackage{physics}
\usepackage{nicefrac}       %
\usepackage{graphicx}
\usepackage{subcaption}

\def\cD{{\mathcal D}}

\def\cF{{\mathcal F}}

\def\cH{{\mathcal H}}

\def\cO{{\mathcal O}}

\def\cT{{\mathcal T}}

\def\cV{{\mathcal V}}

\def\vd{\mathbf{d}}

\newcommand{\mr}[1]{\mathrm{#1}}

\DeclareMathOperator*{\E}{\mathbb{E}}

\DeclareMathOperator*{\argmin}{arg\,min}

\title{DeePKS: a comprehensive data-driven approach towards chemically accurate density functional theory}

\author[a]{Yixiao Chen}
\author[a]{Linfeng Zhang\thanks{linfengz@princeton.edu}}
\author[b]{Han Wang\thanks{wang$\_$han@iapcm.ac.cn}}
\author[a,c]{Weinan E}
\affil[a]{Program in Applied and Computational Mathematics, Princeton University, Princeton, NJ, USA}
\affil[b]{Laboratory of Computational Physics, Institute of Applied Physics and Computational Mathematics, Huayuan Road 6, Beijing 100088, People's Republic of China}
\affil[c]{Department of Mathematics, Princeton University, Princeton, NJ, USA}

\begin{document}

\maketitle

\begin{abstract}
    We propose a general machine learning-based framework for building an accurate and widely-applicable energy functional within the framework of 
    generalized Kohn-Sham density functional theory. 
    To this end, we develop a way of training self-consistent models that are capable of taking large datasets from different systems
    and different kinds of labels.
    We demonstrate that the functional that results from this training procedure
    gives chemically accurate predictions on energy, force, dipole, and electron density 
    for a large class of molecules.
    It can be continuously improved when more and more data are available.
\end{abstract}

\section{Introduction}
Predicting the ground-state information of a many-electron system in an environment of clamped ions is a fundamental task in the field of molecular modeling.
Over the past few decades, a wide variety of methods have been developed for addressing this problem, such as quantum Monte Carlo, post Hartree-Fock (HF) methods (also known as wave function theory, WFT), density functional theory (DFT)~\cite{hohenberg1964inhomogeneous}, etc.
In general, these methods follow a well-known trade-off between accuracy and efficiency. 
The cost of exact WFT methods like full configuration interaction (FCI)~\cite{pople1987QCI} usually scales exponentially with  system size. 
Coupled cluster singles, doubles and perturbative triples (CCSD(T))~\cite{jeziorski1981coupled}, the method often referred to as the golden standard of quantum chemistry, 
has a cost that scales as $\order{N^7}$ with respect to the number of electrons $N$.
The cost of  Kohn-Sham (KS) DFT~\cite{kohn1965self} and its generalized version~\cite{seidl1996generalized} typically scales as  $\order{N^3 \sim N^4}$.
However, currently available DFT models, although much more efficient, are much less accurate compared with FCI and CCSD(T),
due to the approximate nature of the functionals involved.

Developing accurate and efficient DFT functionals is among the world's hardest and most important parameter fitting problems. 
As for all parameter fitting problems, we need a functional form with some free parameters and a way to optimize these parameters.
The key notion in this context is {\it universality}. 
In principle, the DFT functionals are universal and we would like our approximate functionals to be as universal as possible. 
It should  be noted immediately that truly universal and computationally efficient functionals are very difficult, if not impossible, to come by.
Therefore our goal should be  to develop a functional that is efficient and chemically accurate for all the systems that  
can be reasonably represented by the data available.

To this end, we look for models with the following requirements in mind:
\begin{enumerate}
\item We need to have a functional form (for these approximate functionals) that is expressive enough
so that the behavior of different systems, whether small or large molecules or condensed systems, can all be accommodated.
\item We should also  make maximum use of existing high quality data , %
including data for different systems and data with different kinds of labels, such as energy, force, and electron density.
The model should be continuously improvable as more and more data become available.
\end{enumerate}
Since condensed systems involve other non-trivial technical issues, we choose to focus first on molecules. 
For similar reason, we do not discuss the analytical conditions that are used in the so-called non-empirical functionals~\cite{perdew1996generalized,sun2015SCAN}.
Most of these conditions are derived in some limiting cases, such as uniform electron gas.
They are less relevant to molecules or the generalized Kohn-Sham scheme that we are going to use.

A reasonably successful (non-self-consistent) model that accomplishes the first requirement, termed Deep Post Hartree-Fock (DeePHF),  has been developed in Ref.~\citenum{chen2020ground} for molecules.
By exploiting both physical constraints from symmetries and the unprecedented expressivity of neural network (NN) functions,  DeePHF 
succeeded in achieving chemical accuracy for the energy at a cost comparable to Hartree-Fock (HF).   
It has demonstrated impressive performance on existing datasets for molecules.
One main objective of the current work is to extend
DeePHF to a self-consistent framework such as KS-DFT.
We will adopt the  generalized Kohn-Sham (GKS) formalism, with the domain of our functional been elevated from pure density to Kohn-Sham orbitals, so that the  functional space represented is much larger. 
At the same time, we will make sure that  the second requirement listed above is also fulfilled.

Despite several earlier attempts~\cite{snyder2012finding,bogojeski2019density,dick2020machine,lei2019design,liu2017imporving,nagai2020completing}, there have been serious difficulties involved in this task.
For machine learning-based models such as  DeePHF, it was the gradient-based optimization schemes that make them efficiently trainable.
Gradient-based methods can hardly be used in the self-consistent framework, since it is very expensive to compute the gradients of the self-consistent energy, force, and density, with respect to the NN parameters.
For this reason,  an earlier attempt reported in Ref.~\citenum{nagai2020completing} used Monte Carlo, a gradient-free optimization scheme. 
This is prohibitively expensive in the self-consistent setup, particular with large datasets.
When the training data is limited to only the energies of a few molecules,
the pioneering work reported  in Ref.~\citenum{dick2020machine} successfully developed a gradient-based strategy
by effectively decoupling the self-consistent constraint and the gradient-based training.
We will follow a similar strategy, %
but  we have to develop a modified reformulation  to make the process more efficient so that much larger datasets can be handled.
When the training data also include 
alternative  labels other than energy, such as forces and electron density,
to the best of our knowledge, 
no effective gradient-based method has been developed.
We will present a new training scheme that overcomes these difficulties in a very elegant way.

We name the approach proposed here  Deep Kohn-Sham (DeePKS) to highlight the self-consistent nature that distinguishes this method from our previous work. 
Self-consistency enables calculating force and density-related properties naturally from DeePKS, a key feature that differs from a pure energy model.%
We also use DeePKS to refer to the model (i.e. functionals)  obtained this way.
DeePKS obeys all physical and gauge symmetries and is consistent with all known high quality data.
In addition, it can be continuously improved as more and more data become available.
We also note that the training schemes developed here can be used in other situations when 
some self-consistent models are trained.

\section{Methods}

\subsection{(Generalized) Kohn-Sham theory}

We first give a brief overview of the (generalized) Kohn-Sham theory. 
We start from the many-body Schr\"{o}dinger equation of $N$ electrons indexed by $i$, 
\begin{equation}
    \qty( T + W + V_\mr{ext} ) \Psi \qty(x_1,x_2,\dots,x_N) = E_\mr{tot} \Psi \qty(x_1,x_2,\dots,x_N),
\end{equation}
where we use $E_\mr{tot}$ to denote the ground-state energy of the $N$-electron Schr\"{o}dinger equation. 
Here $T = -\frac{1}{2} \nabla^2$ and $W = \frac{1}{2} \sum_{i,j}\frac{1}{|x_i-x_j|}$ denote the kinetic operator and electron-electron interactions, respectively. 
$V_\mr{ext}$ stands for the external potential.%
For example, in an atomic system with $M$ ions indexed by $I$,  $V_\mr{ext} = \sum_i \cV_\mr{ext}(x_i) = \sum_{I,i} \frac{Z_I}{|x_I - x_i|}$.

Following the variational principle, the ground-state energy can also be written as  
\begin{equation}
    E_\mr{tot} 
    = \min_{\Psi} \ev{T+W+V_\mr{ext}}{\Psi}
    = \min_{\Psi} \Bqty\Big{ G_0\qty[\Psi] + E_\mr{ext} \qty[\rho[\Psi]] },
\end{equation}
where%
\begin{align}
    & G_0\qty[\Psi] = \ev{T+W}{\Psi}, \\
    & E_\mr{ext} \qty[\rho] = \int \dd{x} \cV_\mr{ext}(x) \rho(x) .
\end{align}
According to the well-known Hohenberg-Kohn theorem \cite{hohenberg1964inhomogeneous}, this problem is equivalent to another minimization problem with respect to the electron density $\rho$,
\begin{align}
    & E_\mr{tot} = \min_{\rho(x) \rightarrow N} \Bqty\Big{ F_\mr{HK} [\rho] + E_\mr{ext} \qty[\rho]} , 
    \label{eqn:hkmin}\\
    & F_\mr{HK} [\rho] = \min_{\Psi \rightarrow \rho(x)} G_0\qty[\Psi]
    \equiv \min_{\Psi \rightarrow \rho(x)} \ev{T+W}{\Psi}.
    \label{eqn:hkfun}
\end{align}
Eq.~\ref{eqn:hkfun} defines the Hohenberg-Kohn (HK) functional $F_\mr{HK} [\rho]$ using the Levy-Lieb constrained search formulation \cite{levy1979universal,lieb1983density}. Note here both $F_\mr{HK} [\rho]$ and $ G_0\qty[\Psi]$ are considered to be universal, meaning that they do not depend explicitly on the external potential $V_\mr{ext}$.

Directly solving the ground-state energy or representing the HK functional can be very difficult, since it involves dealing with the $N$-particle wave function. Therefore, one often resorts to the popular Kohn-Sham (KS) scheme to simplify this problem. 
The key ingredient of KS-like theories is to replace the general $N$-particle ground-state $\Psi$ with a model system, whose ground state can be represented by a single Slater determinant $\Phi = \frac{1}{\sqrt{N!}}\mr{det}\qty[\varphi_i\qty(x_j)]$, where we use $\qty{\varphi_i}$ to denote a set of orthonormal single particle orbitals.  
The  energy functional can also be written as  $G\qty[\Phi] = G\qty[\Bqty{\varphi_i}]$.  %
As a result, the ground-state energy $E_\mr{KS}$ and density functional $F_\mr{KS}$  is given by
 \begin{align}
    & E_\mr{KS} = \min_{\rho(x) \rightarrow N} \Bqty\Big{ F_\mr{KS} [\rho] + E_\mr{ext} \qty[\rho]} , 
    \label{eqn:ksmin}\\
    & F_\mr{KS} [\rho] = \min_{\Phi \rightarrow \rho(x)} G\qty[\Phi] 
    = \min_{ \substack{
    \Bqty{\varphi_i}\rightarrow \rho(x) 
    \\ 
            \braket{\varphi_i}{\varphi_j} = \delta_{ij} }} 
      G\qty[\Bqty{\varphi_i}]. 
    \label{eqn:ksfun}
\end{align}

Depending on how the  functional $G\qty[\Phi]$ is chosen, the above formulation gives many different theories. To name a few:
\begin{itemize}
    \item If we leave $G$ unchanged from $G_0$, we get the Hartree-Fock theory,
    \begin{equation}
    \label{eqn:hfdef}
        G_\mr{HF}\qty[\Phi] = G_0\qty[\Phi] = \ev{T+W}{\Phi} = \ev{T}{\Phi} + E_H[\rho] + E_F[\Bqty{\varphi_i}],
    \end{equation}
    where $ E_H[\rho] $ and $ E_F[\Bqty{\varphi_i}] $ denote the Coulomb (Hartree) and exchange (Fock) energy, respectively. 
    Note here $E_H$ depends only on the electron density $\rho(x) = \sum_i \abs{ \varphi_i(x) }^2$.
    \item If we constrain $G$ such that the only term that explicitly depends on $\Phi$ is the kinetic energy, we get the standard KS theory,
    \begin{equation}
    \label{eqn:sksdef}
        G_\mr{KS}\qty[\Phi] = \ev{T}{\Phi} + E_H[\rho] + E_{xc}[\rho],
    \end{equation}
    where $ E_{xc}[\rho] $ is the so called exchange-correlation functional.
    Usually $ E_{xc}[\rho] $ can be split into two parts, the exchange energy $ E_{x}[\rho] $  and the correlation energy $ E_{c}[\rho] $ . 
    \item If we include part of the Fock exchange operator in addition to the standard exchange-correlation functional, we get a standard version the hybrid Kohn-Sham theory,
    \begin{equation}
        G_\mr{Hyb}\qty[\Phi] = \ev{T}{\Phi} + E_H[\rho] + \lambda E_F[\Bqty{\varphi_i}] + (1-\lambda) E_{x}[\rho] + E_{c}[\rho],
    \end{equation}
    where $\lambda$ is a tunable factor deciding how much the exact exchange operator is used.
\end{itemize}
The term {\it generalized} Kohn-Sham (GKS) theory simply refers to any choice of $G$ that does not satisfy the standard KS condition (Eq.~\ref{eqn:sksdef}). Many functionals fall into this class, including all  hybrid functionals and most Meta-GGA functionals.

A KS-like theory is considered to be exact if its choice of $G$ yields the same density functional as the original Hohenberg-Kohn functional, namely, 
\begin{equation}
     F_\mr{KS} [\rho] = F_\mr{HK} [\rho].
\end{equation}
Therefore, an exact theory would give the exact ground-state energy, $E_\mr{KS} = E_\mr{tot}$, as well as the exact ground-state density $\rho$. 
As an example, the aforementioned Hartree-Fock theory is obviously not exact. 
It remains an open question whether there exist a possible choice of $G$ in general that yields the exact functional, and hence the exact ground-state density. 
In the context of standard KS theory, it is termed the problem of non-interacting $v$-representability.
From this point of view, the GKS theory is at least as exact as the standard KS theory.

In order to solve the KS-like problem, we reformulate Eq.~\ref{eqn:ksmin} as a direct minimization problem with respect to the single particle orbitals $\Bqty{\varphi_i}$, namely
\begin{equation}
\label{eqn:emin}
\begin{aligned}
    E_\mr{KS} 
      = \min_{\Bqty{\varphi_i},\braket{\varphi_i}{\varphi_j} = \delta_{ij}}  
        \Bqty\Big{  G\qty[\Bqty{\varphi_i}]  +  E_\mr{ext} \qty[\rho[\Bqty{\varphi_i}]]}.
\end{aligned}
\end{equation}
We now further require that the functional derivative of $ G\qty[\Bqty{\varphi_i}] $ can be cast into the form of a single particle operator,
\begin{equation}
    \fdv{G\qty[\Bqty{\varphi_j}]}{\bra{\varphi_i}} = \cO[\Bqty{\varphi_j}] \ket{\varphi_i}.
\end{equation}
Therefore, using Lagrange multipliers on Eq.~\ref{eqn:emin}, we obtain the self consistent field (SCF) equation:
\begin{equation}
    \cH[\Bqty{\varphi_j}] \ket{\varphi_i}
    \equiv \qty(\cO[\Bqty{\varphi_j}] + \cV_\mr{ext}) \ket{\varphi_i} 
    = \varepsilon_i \ket{\varphi_i} \qfor i = 1 \dots N.
    \label{eqn:scf}
\end{equation}
where we use $\cH$ to denote the single particle Hamiltonian.
As an example, for the HF theory (Eq.~\ref{eqn:hfdef}),   we have
\begin{equation}
    \cO_\mr{HF}[\Bqty{\varphi_j}] = \cT + \cV_H[\rho] + \cV_F[\Bqty{\varphi_j}].
\end{equation}
For the standard KS theory (Eq.~\ref{eqn:sksdef}),  we have
\begin{equation}
    \cO_\mr{KS}[\Bqty{\varphi_j}] = \cT + \cV_H[\rho] + \cV_{xc}[\rho].
\end{equation}
Here we use $\cT$, $\cV_H$, $\cV_F$, $\cV_{xc}$ to denote single particle kinetic, Coulomb, exact exchange and exchange-correlation operators, respectively.

\subsection{Model Construction}
We construct our GKS model on top of an existing KS-like model and add a parametrized correction term $E_\delta$ to it. 
To be more specific, we define our  energy functional to be 
\begin{equation}
    G\qty\big[\Bqty{\varphi_i} | \omega] = G_\mr{base}[\Bqty{\varphi_i}] + E_\delta\qty\big[\Bqty{\varphi_i} | \omega]
\end{equation}
where $\omega$ stands for the set of parameters we use in the representation of $E_\delta$. 
The corresponding single particle Hamiltonian is then given by
\begin{equation}
    \cH\qty\big[\Bqty{\varphi_i} | \omega] = \cO_\mr{base}[\Bqty{\varphi_j}] + \cV_\mr{ext} + \cV_\delta\qty\big[\Bqty{\varphi_i} | \omega].
\end{equation}
The reference point $G_\mr{base}$ should be a reasonable electron energy functional in KS-like theories, e.g., $G_\mr{HF}$, $G_\mr{KS;PBE}$, $G_\mr{Hyb;SCAN0}$, etc.

Before proceeding further, we list the set of requirements that we ideally want   $E_\delta\qty\big[\Bqty{\varphi_i} | \omega]$ to obey: 
1) {\it Generality}. The model should be general enough to be applicable for all the systems whose local electronic configurations are well represented by the training data.
2) {\it Locality}. The model should be relatively local, so that it can potentially be constructed using data from small systems and then be generalizable to larger ones.
3) {\it Symmetry}. The model should respect both physical and gauge symmetries.
Here physical symmetry means that $E_c$ should be invariant under translation and rotation of the system. %
Gauge symmetry means that $E_\delta$ should be invariant when the occupied orbitals $\{\ket{\varphi_i}\}$ undergo a unitary transformation.
4) {\it Accuracy}. For target systems, the model should achieve chemical accuracy, i.e.~a prediction error lower than 1 kcal/mol.
5) {\it Efficiency}. The cost for solving the model should be comparable to that of HF or other DFT models.

To satisfy these requirements, we follow our previous work\cite{chen2020ground} to construct $E_\delta$ as a neural network model using the ``local density matrix'' as input. 
Briefly speaking, we build our functional based on the one-particle reduced density matrix
\begin{equation}
\Gamma(x,x') = \sum_i \bra{x}\ket{\varphi_i}\!\bra{\varphi_i}\ket{x'} = \sum_i \varphi_i^*(x') \varphi_i(x).
\end{equation}
We then project it onto a set of atomic basis $\Bqty{\alpha^I_{nlm}}$ indexed by the radial number $n$, azimuthal number $l$, magnetic (angular) number $m$ and centered on each atom $I$, to get the ``local density matrix''
\begin{equation}
    \qty(\cD^I_{nl})_{mm'} = \sum_i \braket{\alpha^I_{nlm}}{\varphi_i}\!\braket{\varphi_i}{\alpha^{I}_{nlm'}}.
\end{equation}
Note here for simplicity and locality, we only take the block diagonal part of the full matrix, i.e. indices $I$, $n$ and $l$ are taken to be the same for both sides of the projection, only angular indices $m$ and $m'$ differ. For fast overlap evaluation, we use standard GTO functions but with customized coefficients to make the basis set complete enough. A total of 108 basis functions is used for each atom. The detailed coefficients can be found in the appendix of Ref.~\citenum{chen2020ground}.

To deal with the rotational symmetry of the basis  $\Bqty{\alpha^I_{nlm}}$, we use the eigenvalues of the local density matrix as our descriptor 
\begin{equation}
    \vd^I_{nl} = \mr{EigenVals}_{mm'} \left[\left(\cD^I_{nl}\right)_{mm'}\right],
\end{equation}
and we use a neural network model to output the ``correction'' energy
\begin{equation}
    E_\delta = \sum_I \cF^\mr{NN} \qty(\vd^I).
\end{equation}
Hence the corresponding potential $\cV_\delta$ is given by
\begin{equation}
    \label{eq:vc}
    \cV_\delta = \sum_{Inlmm'}\pdv{E_\delta}{\left(\cD^I_{nl}\right)_{mm'}}
                              \ketbra{\alpha^I_{nlm}}{\alpha^I_{nlm'}}.
\end{equation}

We emphasize that although $E_\delta$ is constructed from the one particle density matrix, neither the ground-state orbitals nor the density matrix calculated by our model should be expected to have a physical meaning.
Instead, we consider the ground-state density to be physical, just as in the standard KS theory, and expect it to coincide with the true ground-state density once we have the exact functional.
This is because we follow the GKS approach, rather than a 1-reduced density matrix functional theory~\cite{gilbert1975hohenberg}, which can not be mapped to a KS system.

\subsection{Training Algorithms}
We now discuss how to train a self-consistent model. 
Here self-consistency means that the property predicted by the model 
is obtained via
a minimization process 
and is given at the minimum. 
A KS-like DFT method is naturally self-consistent. 
On the contrary, methods like Møller–Plesset perturbation theory\cite{moller1934note} and many other post-HF theories are not self-consistent, since they do not involve a minimizing procedure. 
We call those methods energy models, to be distinguished from the self-consistent ones.
In this context, recent machine learning-based schemes, such as  DeePHF method~\cite{chen2020ground} and the MOB-ML method~\cite{cheng2019universal}, are energy models.

Similar to other supervised learning procedures, We fit the energy functional using existing datasets with certain labels. 
These labels can be acquired from calculations of high-accuracy methods, such as CCSD(T) and quantum Monte Carlo. 
Generally speaking, we consider three types of labels: %
\begin{enumerate}
    \item quantity that is the direct output of the functional after a minimization procedure. Here it is the total energy.
    \item quantity that depends on both the direct output of the functional and its minimizer.
    Here we consider the atomic force. 
    \item quantity that depends on the minimizer of the functional, but only implicitly through the mathematical form of the functional.
    Here we consider the ground-state density.
\end{enumerate}
As has been mentioned, using all these labels in training is a non-trivial task, since there is a highly complicated and expensive procedure for calculating the corresponding quantities.  %
Here we develop  general and efficient training algorithms for these three types of labels. 

\textbf{Type one (energy)}. The training procedure with the energy label may seem straightforward at first glance.
Using the $\ell^2$ norm as the error metric, the optimization problem becomes
\begin{equation}
\label{eqn:direct_loss}
    \min_\omega  \E_\mr{data} \left[ \left(E_\mr{label} - \min_{\{\varphi_i\}, \braket{\varphi_i}{\varphi_j} = \delta_{ij} } E_\mr{model} \qty\big[\{\varphi_i\} | \omega ] \right)^2 \right].
\end{equation}
Here 
\begin{equation}
    E_\mr{model} \qty\big[\{\varphi_i\} | \omega ] = G_\mr{base} [\Bqty{\varphi_i}] + E_\mr{ext} [\rho[\Bqty{\varphi_i}]] + E_\delta \qty\big[\{\varphi_i\} | \omega ],
\end{equation}
where %
the expectation is taken over the training samples. 

The gradient of $E_\mr{model}$ with respect to $\omega$ can be easily obtained using the Hellmann-Feynman theorem.
However, the minimization procedure of  $\{\varphi_i\}$ involves solving an SCF equation (Eq.~\ref{eqn:scf}) that is very time consuming. 
A typical training procedure consists of as many as a million gradient descent steps. 
This is unrealistic if the SCF equation is solved at every step.

We use a different optimization formalism.
Instead of treating the minimized energy as a function of the parameters $\omega$, we consider it as a function of both orbitals $\{\varphi_i\}$ and parameters $\omega$ that satisfies the constraint that $\{\varphi_i\}$ is the minimizer. 
Therefore, the whole optimization problem can be written as
\begin{align}
    \min_\omega \quad & \E_\mr{data} \left[  \left(E_\mr{label} -  E_\mr{model} \qty\big[\{\varphi_i\} | \omega ] \right)^2 \right] \label{eqn:iter_loss} \\
    \mr{s.t.}\quad & \ \exists\, \varepsilon_i \leq \mu 
                    \qc \left(\cH \qty\big[\{\varphi_i\} |\omega]- \varepsilon_i \right) \ket{\varphi_i} = 0
                    \qc \braket{\varphi_i}{\varphi_j} = \delta_{ij} 
                    \qfor i,j = 1 \dots N \label{eqn:iter_cons} 
\end{align}
where 
Eq.~\ref{eqn:iter_cons} is a parameterized version of Eq.~\ref{eqn:scf}, i.e., 
the single particle Hamiltonian $\cH \qty\big[\{\varphi_i\} |\omega]$ depends on both the orbitals $\qty{\varphi_i}$ and the model parameters $\omega$. Here $\mu$ is the chemical potential and $\varepsilon_1 \leq \varepsilon_2 \leq \dots \leq \varepsilon_N$ denote the lowest $N$ eigenvalues.

\begin{figure}[htb]
    \centering
    \includegraphics[width=0.6\textwidth]{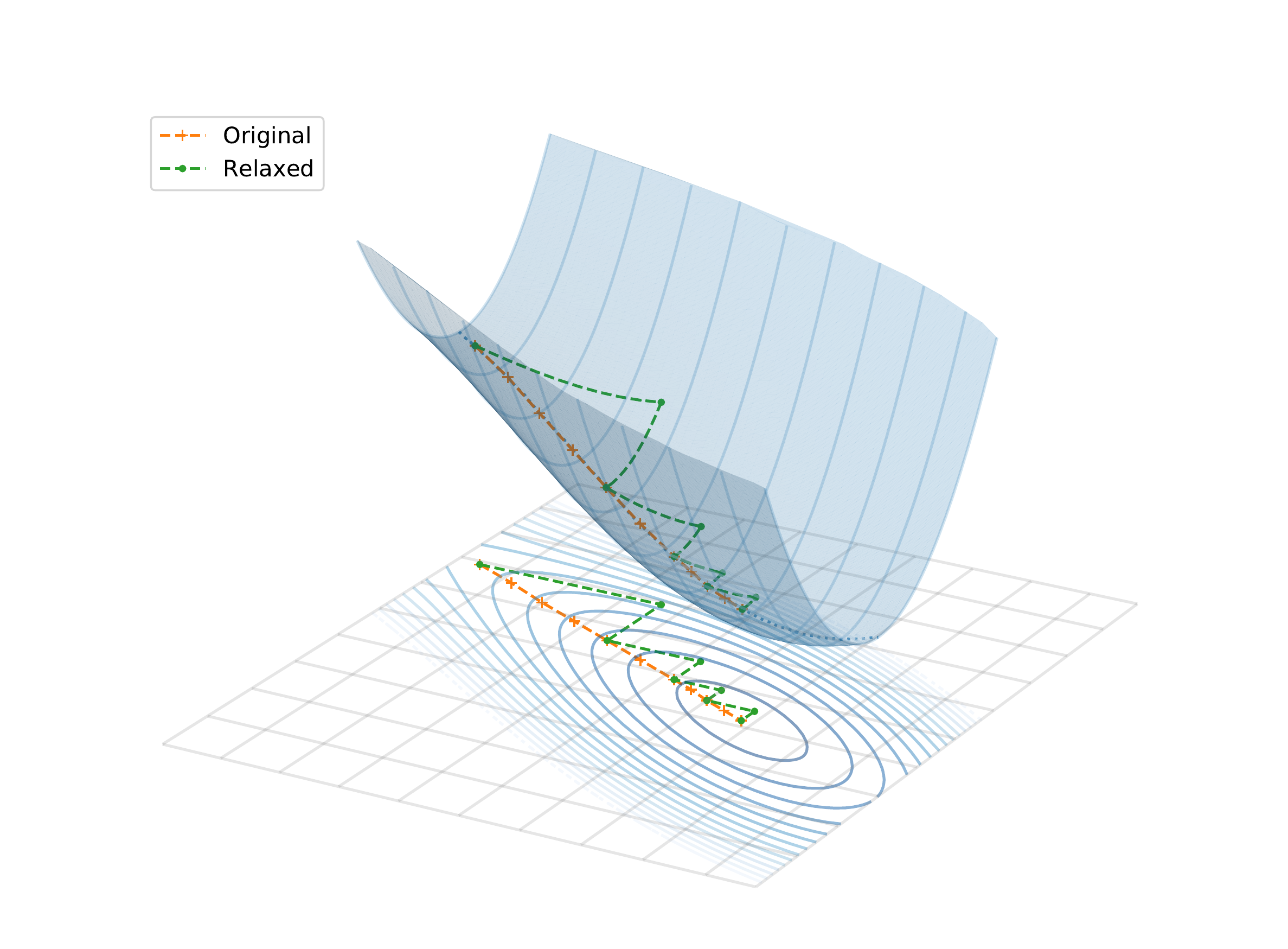}
    \caption{A schematic illustration of the iterative training procedure using the projection method. Here ``Original'' stands for the direct constrained minimization following Eq.~\ref{eqn:direct_loss} and ``Relaxed'' stands for the relaxed projection method in Eq.~\ref{eqn:iter_loss}-\ref{eqn:iter_cons}. In step 2 of the relaxed method, the optimization of $\omega$ strays away from the minimizing manifold, while in step 3 the projection of $\Bqty{\varphi_i}$ brings it back.} 
    \label{fig:optim}
\end{figure}

We now can use a projection method to relax the constraint and this  reduces  the cost of calculating the SCF equation. 
In other words, we can first optimize the parameters $\omega$ using unconstrained gradient-based method with the orbitals $\{\varphi_i\}$ fixed. After several steps, we project the orbitals back to the constraint manifold by solving the SCF equation. Decreasing the projection frequency can largely reduce the computational cost since most of the computation time is spent in the SCF equation. 
To make it more clear, we write the procedure into the following steps.
\begin{enumerate}
    \item Initialize a set of $\qty{\varphi_i}$ and $\omega$ that satisfies the SCF equation, e.g., take $\omega$ to be all zero and $\qty{\varphi_i}$ to be the Hartree-Fock solution. Also keep track of the predicted energy $E_\mr{model}$.
    \item Update the parameters $\omega$ by training the model following Eq.~\ref{eqn:iter_loss} with fixed orbitals $\{\varphi_i\}$.  
    \item Update the orbitals $\{\varphi_i\}$ by solving the SCF equation with fixed model parameters $\omega$.
    \item Check whether the predicted energy $E_\mr{model}$ converges. If not, go to step 2 and do more iterations.
\end{enumerate}
A schematic illustration of this approach is shown in Fig.~\ref{fig:optim}. 
Note that  we usually take many training steps in step 2. %
In practice, when restarting from old parameters using new orbitals, we find it possible to train the model until the validation error no longer decreases, without breaking the convergence of the whole procedure. 
Therefore, the total time of solving SCF equation is significantly reduced. 

We note that a similar formalism has been proposed and used by the NeuralXC scheme~\cite{dick2020machine}.
The major difference is that, in DeePKS, a single NN function is used as a universal approximator.
The function form does not change with the iterative process, and its parameters does not depend on the chemical species of the associated atom.
In contrast, in NeuralXC, the parameters depend on the chemical species, and in each iteration, a new NN layer is appended to the NN model from the previous iteration.
The reformulation in DeePKS  is designed to makes it more transferable to  larger chemical space, and more suited for  larger dataset.

\textbf{Type two (force)}. The atomic forces from the proposed model can be easily calculated by the standard Hellmann-Feynman theorem,
\begin{equation}
\begin{aligned}
    F_\mr{model} \qty\big[\{\varphi_i^*[\omega]\} |\omega] &= -\pdv{E_\mr{model} \qty\big[\{\varphi_i^*\} |\omega]}{X} \\
                 &= F_\mr{base} \qty\big[\{\varphi_i^*\}]
                 - \sum_{Inlmm'}\pdv{E_\delta \qty\big[\{\varphi_i^*\} |\omega]}{\left(\cD^I_{nl}\right)_{mm'}} 
                   \sum_i \ev**{ \pdv{ \left( \ketbra{\alpha^I_{nlm}} \right) }{X} }{\varphi_i^*},
\end{aligned}
\end{equation}
where we have written out the dependence on the parameters $\omega$ explicitly. 
We use $\{\varphi_i^*\}$ to denote the minimizer of the total energy functional, which themselves are functions of $\omega$, 
\begin{equation}
    \{\varphi_i^*[\omega]\} = \argmin_{\braket{\varphi_i}{\varphi_j} = \delta_{ij} } E_\mr{model} \qty\big[\{\varphi_i\} |\omega].
\end{equation}

We can see that the force $F$ depends directly on both the model parameters $\omega$ and the  minimizing orbitals $\{\varphi_i^*[\omega]\}$.
This introduces an additional difficulty when we evaluate the gradient of $F$ with respect to $\omega$. 
The contribution from the $\{\varphi_i^*[\omega]\}$ term is very hard to compute, since it involves a whole minimization procedure, and there is no Hellmann-Feynman theorem to save us. 

Luckily, this difficulty disappears in our iterative training procedure, where the gradient we used to optimize $\omega$ is no longer the constrained one. The orbitals are treated as independent variables so that they do not contribute to the gradient. 
Therefore, the gradient can be calculated straightforwardly using a back propagation procedure. 
By writing the force term into the loss function,
the new optimization problem becomes
\begin{equation}
\label{eqn:forcetrain}
\begin{aligned}
    \min_\omega \quad & \E_\mr{data} \left[  \left(E_\mr{label} -  E_\mr{model} \qty\big[\{\varphi_i\} | \omega ] \right)^2 + \lambda_f \left(F_\mr{label} -  F_\mr{model} \qty\big[\{\varphi_i\} | \omega ] \right)^2 \right] \\
    \mr{s.t.}\quad & \ \exists\, \varepsilon_i \leq \mu 
                    \qc \left(\cH \qty\big[\{\varphi_i\} |\omega]- \varepsilon_i \right) \ket{\varphi_i} = 0
                    \qc \braket{\varphi_i}{\varphi_j} = \delta_{ij} 
                    \qfor i,j = 1 \dots N ,
\end{aligned}
\end{equation}
where $\lambda_f$ is a tunable parameter that  determines the weight of force label in the loss function. 
We can then use the iterative algorithms described above to solve this optimization problem. %

\textbf{Type three (density)}. The ground-state density given by the proposed model is a function of the minimizing orbitals,
\begin{equation}
    \rho_\mr{model}(x) = \sum_i^N |\varphi_i^*(x)|^2.
\end{equation}
Since it does not depend on the parameters $\omega$ explicitly, unlike the case for forces, we cannot write the density into the loss function.
To solve this problem,  we introduce a penalty term in the SCF equation to ``guide'' the training procedure. 
This is done by  changing the minimization problem in Eq.~\ref{eqn:emin} into:
\begin{equation}
\label{eqn:emin_dens}
     \min_{\{\varphi_i\}, \braket{\varphi_i}{\varphi_j} = \delta_{ij} } 
     \Bqty\Big{ E_\mr{model} \qty\big[\{\varphi_i\} | \omega ] + \lambda_\rho D\qty\big[\rho[\{\varphi_i\}], \rho_\mr{label}] },
\end{equation}
where $\lambda_\rho>0$ is the strength of the penalty and $D$ is some non-negative error metric that equals to zero only when $\rho = \rho_\mr{label}$. 
Hence, if the SCF solution gives the exact density, the penalty term does not influence the minimizer. 
Otherwise, 
according to the discussion in the Appendix,
because of an additional potential term in the SCF equation,
\begin{equation}
    \cV_\mr{pnt}[\rho | \rho_\mr{label}] = \fdv{ D[\rho, \rho_\mr{label}]}{\rho},
\end{equation}
it will lead to a self-consistent energy strictly larger than the one obtained without this penalty term, and a density that is closer to the label.

Note here $\lambda_\rho$ does not need to be a fixed value. Rather, it can be a bunch of values or even a non-negative random variable. When the model yields exact density, all the functionals with different $\lambda_\rho$ should give the exact solution. 
When the solution is not exact, randomized $\lambda_\rho$ services as a regulator that helps reducing the overfitting and provides better results compared to using a single fixed value, and is more efficient than using multiple values. 
If we choose it to be a random variable, the modified optimization problem becomes:
\begin{equation}
\label{eqn:denstrain}
\begin{split}
    \min_\omega \quad & \E_{\mr{data}, \lambda_\rho} \left[  \left(E_\mr{label} -  E_\mr{model} \qty\big[\{\varphi_i\} | \omega ] \right)^2 + \lambda_f \left(F_\mr{label} -  F_\mr{model} \qty\big[\{\varphi_i\} | \omega ] \right)^2 \right] \\
    \mr{s.t.}\quad & \ \exists\, \varepsilon_i \leq \mu 
                    \qc \left(\cH \qty\big[\{\varphi_i\} |\omega] 
                        + \lambda_\rho \cV_\mr{pnt} \qty\big[\rho[\{\varphi_i\}] | \rho_\mr{label}] 
                        - \varepsilon_i \right) \ket{\varphi_i} = 0
                    \qc \braket{\varphi_i}{\varphi_j} = \delta_{ij} \\
                   & \hskip 0.55\textwidth \qfor i,j = 1 \dots N
\end{split}
\end{equation}
The same projection-based training procedure  can be applied to this loss function.

We note that although in this paper we take energy, force, and density as examples, these algorithms are rather general and can be easily transferred to similar learning problems that involve an optimization procedure for the evaluation of meaningful quantities. 
For example, if we include dipole as label, we can add a penalty term similar to Eq.~\ref{eqn:emin_dens}.
Moreover, the training algorithms are not limited to the specific GKS model we described above. 
Instead, they can be applied to gradient-based optimization tasks for any exchange correlation functionals and even other self-consistent learning problems.

\subsection{Related Works}
{Before reporting numerical results, we discuss a few related work to the DeePKS scheme, in the spirit of developing machine learning assisted physical models.
First, there have been some efforts on using deep neural networks to parameterize the many-electron-ion trial wavefunction, and using a variational Monte Carlo (VMC) approach to optimize the parameters. 
The first attempt was reported by Ref.~\citenum{han2019solving}. This is followed by some more recent efforts~\cite{hermann2020deep, pfau2020ab}. The purpose of these efforts is to solve the original quantum many-body electron problem.
In comparison,  DeePKS takes results from the quantum many-body electron problem as inputs and attempts to 
parametrize the exchange-correlation functionals.

Secondly, there have been some efforts on using machine learning based schemes to represent quantities that are functions of atomic positions and their chemical species. 
An incomplete list includes Refs.~\citenum{behler2007generalized, bartok2010gaussian, rupp2012fast, ramakrishnan2015big, chmiela2017machine, schutt2017schnet, smith2017ani, han2017deep, zhang2018deep, zhang2018end, brockherde2017bypassing, grisafi2018transferable, chandrasekaran2019solving, zepeda2019deep}.
In particular, Ref.~\citenum{rupp2012fast} reports a kernel-based method for fast and accurate modeling of molecular atomization energies;
Ref.~\citenum{ramakrishnan2015big} reports a $\Delta$-learning approach, which shares a similar spirit of our work in a different context. Such an idea of fitting the difference between a baseline model and target values has been widely adopted by the machine learning community, see, for example, the Gradient Boosting Machine\cite{friedman2001greedy} that iterates the delta fitting procedure for multiple times in a more systematic way.} %

\section{Results}
We now examine the performance of the DeePKS scheme on three classes of data that have been used for benchmark purposes in the literature. %
Unless otherwise specified, all labels are given by  CCSD(T), and all calculations are conducted using the cc-pVDZ basis. 
\begin{itemize}
    \item Malonaldehyde, including 1500 configurations with energy, force, and density labels. 
    We use this dataset to test thoroughly our training method with all  three types of labels. 
    Since within the CCSD(T) formalism, perturbative triple does not give the corresponding density, we use CCSD  for density related tests. 
    The data is calculated from PySCF\cite{sun2018pyscf} with molecular configurations coming from the  sGDML dataset\cite{sauceda2019molecular}.
    \item Three molecules (malonaldehyde, benzene and toluene), including 1500 configurations for each molecule with energy and force labels. This is a subset of the sGDML dataset\cite{sauceda2019molecular} under the same numerical setup, therefore we can train one model on all three molecules 
    and examine the inter-molecule performance, as a first step toward universal functionals.
    \item QM7b-T dataset\cite{cheng2019data}, including 7212 molecules and one configuration for each molecule, with energy labels only. 
    This  is the largest publiclly available dataset with CCSD(T) accuracy. 
    It has been used to benchmark several other methods \cite{cheng2019universal,cheng2019regression,christensen2020fchl} as well as  the energy model developed in \cite{chen2020ground}. 
    We test it here to make a comparison of the new self-consistent model and the previous energy model.
    We also use it  to examine the ability of the DeePKS method for generating  ``universal'' functionals that are applicable to as many systems  as possible.
\end{itemize}
We emphasize  that the objective of our method is to build one single functional with chemical accuracy for as many systems as possible, although it is currently limited by the data we have. 
The functional should be able to predict accurate results for all the systems that are well represented in the training set, and its coverage can be enlarged continuously by adding more and more training data.

We implement the DeePKS method using the open-source packages PySCF\cite{sun2018pyscf} and PyTorch\cite{PyTorch2019}. 
We start our iteration from a functional obtained using DeePHF and orbitals solved from that functional. 
In each iteration, the optimization of the neural network parameters is conducted for 10,000 epochs using the ADAM optimizer\cite{kingma2014adam}. 
For all training that includes force labels, we set the parameter $\lambda_f$ to be 0.1. 
One more trick we use is that, to speed up the convergence, after training with ADAM, we further correct the model with a global energy shift, which is calibrated from the training set.

We now examine the performance our method on the malonaldehyde molecule, using the HF functional as the base model, $G_\mr{base} = G_\mr{HF}$.
As a first step, to have an intuitive picture of the newly proposed iterative method,
we use energy and force as training labels and study the behavior of the mean absolute error (MAE) in the testing set during the training process. 
The error for the forces is calculated component-wise.
The training is done on 1000 molecular configurations and testing on the remaining 500. 
We also include result from sGDML\cite{sauceda2019molecular} and DeePMD model \cite{zhang2018deep,zhang2018end} for comparison.

\begin{figure}[htb]
    \centering
    \includegraphics[width=0.6\textwidth]{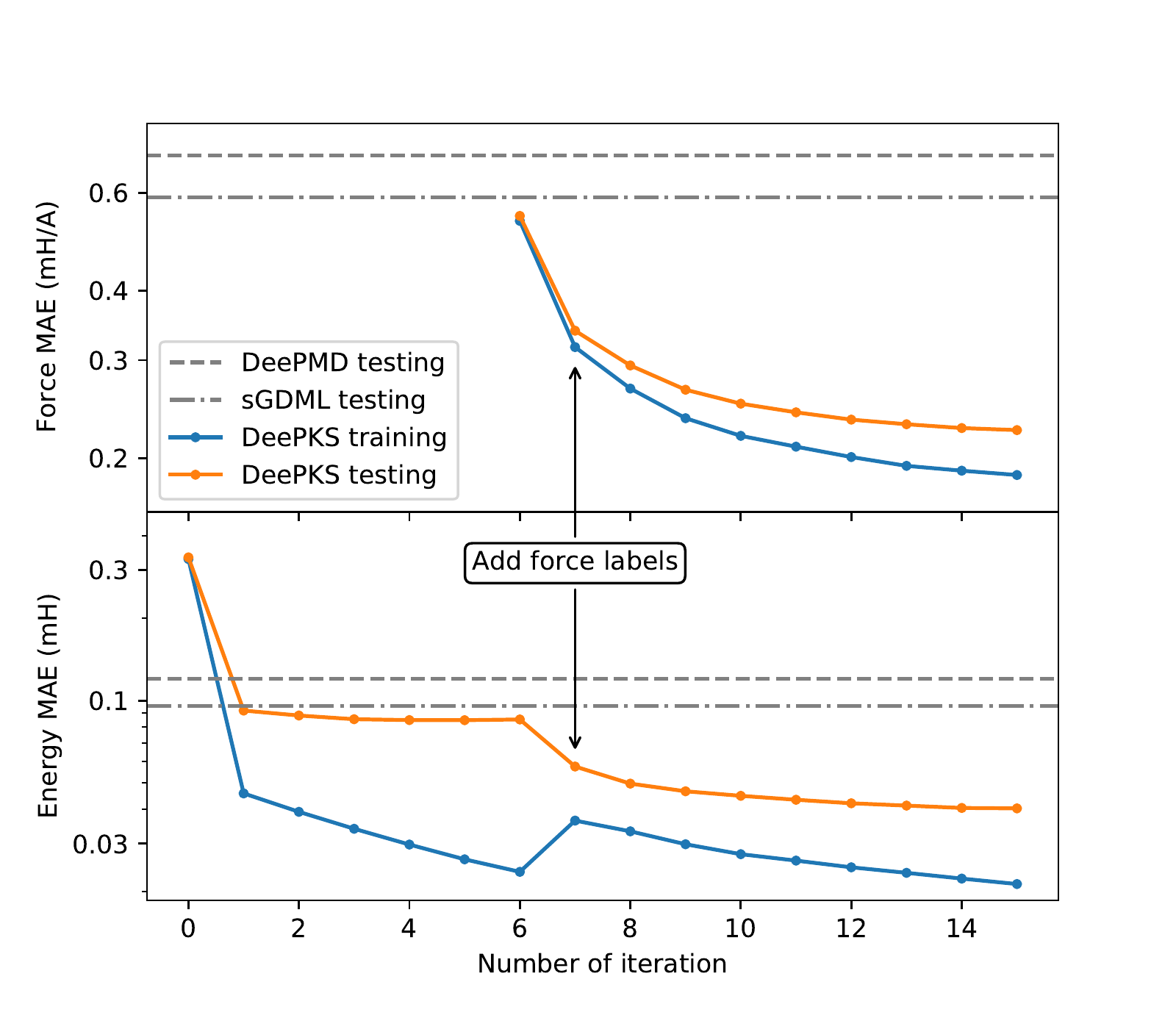}
    \caption{The energy and force errors during the training process for the malonaldehyde  dataset. Force labels are added starting from iteration 7. Results from DeePMD and sGDML are included for comparison.}
    \label{fig:mda_train}
\end{figure}

As can be seen in Fig.~\ref{fig:mda_train}, when training with only energy labels (iterations 0 to 6), the testing accuracy quickly saturates while the training error keeps decreasing, suggesting that the model begins to overfit. %
On the other hand, even though we  train with only energy labels, the model already outperforms DeePMD and sGDML methods, both of which utilize forces as training labels. 
When we include force labels after iteration 6, the testing accuracy can be further improved by two to three times.
This shows  the effectiveness of adding force labels in the training.

\begin{figure}[htb]
    \centering
    \includegraphics[width=0.6\textwidth]{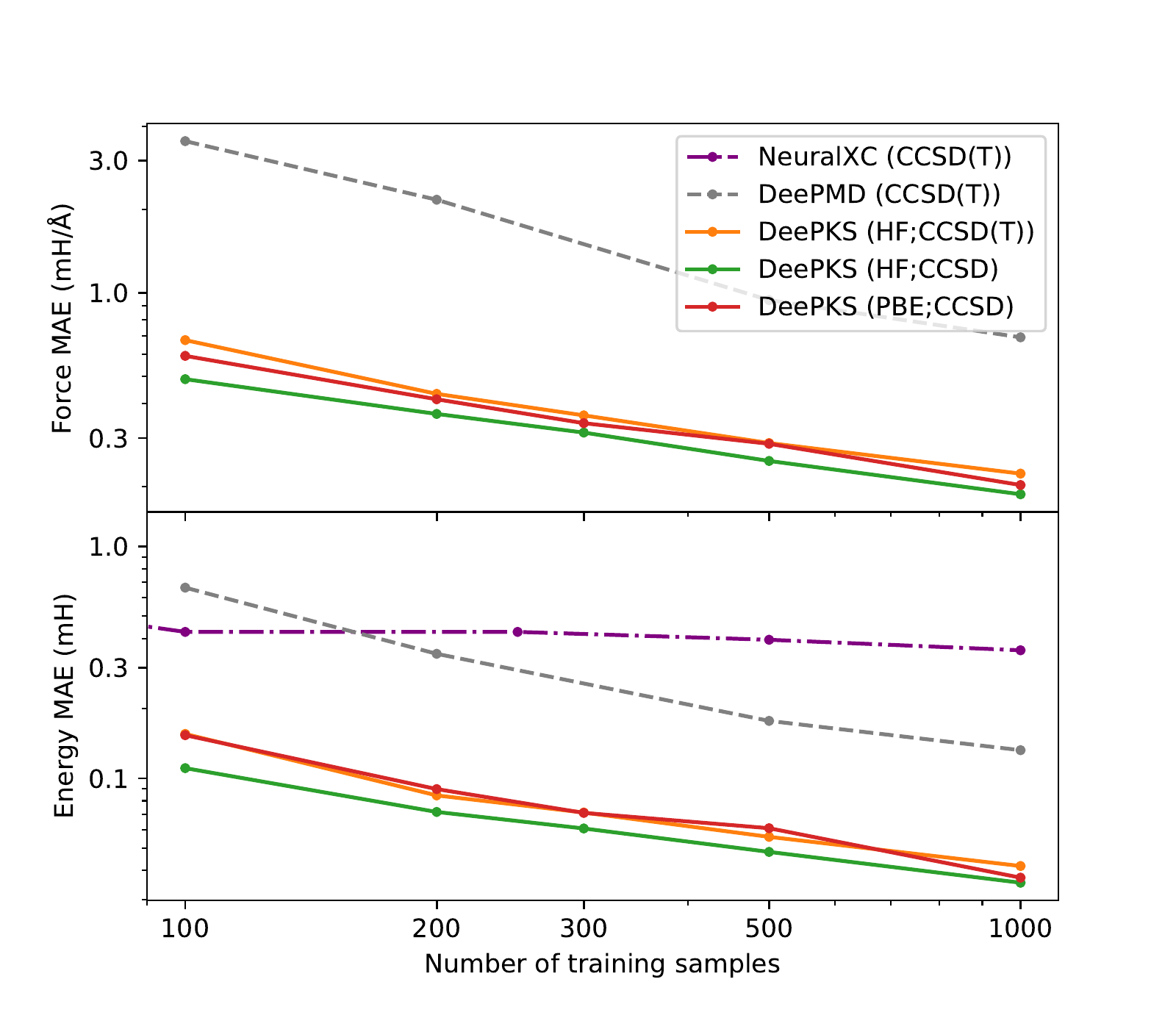}
    \caption{The learning curve of both energy and force for the malonaldehyde  dataset. Results from DeePMD and NeuralXC are included for comparison. NeuralXC results are digitally captured from Ref.~\citenum{dick2020machine}.}
    \label{fig:mda_lc}
\end{figure}

To further examine the sample efficiency of our method, we study the learning curve associated with the malonaldehyde molecule by plotting the testing MAE of both energies and forces versus the number of training samples.
Each time the dataset is augmented, existing samples in the dataset are kept, and the testing error is calculated on the rest part of the data. 
For comparison, we include the result of NeuralXC\cite{dick2020machine} and DeePMD. %
As shown in Fig.~\ref{fig:mda_lc}, in all cases, DeePKS  outperforms both DeePMD and NeuralXC: %
Using the same amount of training data, the accuracy of both the energies and forces is improved 3 to 10 times. 
As an ablation study, we also examine the situation of using labels at the CCSD level and starting from PBE functionals\cite{perdew1996generalized} ($G_\mr{base} = G_\mr{KS;PBE}$),
we find that the results do not change much.  %
Therefore, hereafter  we focus on the HF based model, since the implementation of PBE in PySCF is rather slow. 

We now move to density related tests. 
Here we use labels at the CCSD level. 
We follow Eq.~\ref{eqn:denstrain} to train our model with density labels. 
The error penalty term is taken to be the Coulomb repulsion energy of the density difference,
\begin{equation}
\begin{aligned}
    D[\rho, \rho_\mr{label}] &= \int \dd{x_1} \dd{x_2} \frac{\Delta \rho(x_1)\Delta \rho(x_2)}{\qty|x_1-x_2|} \\
    \Delta \rho(x) &=  \rho(x) - \rho_\mr{label}(x),
\end{aligned}
\end{equation}
which can be evaluated with very small cost in PySCF. 
The penalty parameter $\lambda_\rho$ is sampled uniformly from 0 to 1 for every data point and every SCF calculation. 
We train with this setup for 20 iterations and then remove the penalty and perform another 5 iterations for relaxation. 
As we will see later, such relaxation will slightly reduce the accuracy for density, but substantially improve the accuracy for energy and force.

We study the performance of the DeePKS model in terms of the prediction error of energy $E$, force $F$, dipole $\mu$ and point-wise electron density $\rho$. 
For comparison, we also include different training schemes and results from several other methods.
We use the $\ell^1$ norm for energy and density, the component-wise $\ell^1$ norm for force and $\ell^2$ norm for dipole as error metrics. 
All models are trained on 1000 malonaldehyde configurations and the testing errors are averaged over the rest 500 configurations. 
For HF and DFT functionals, a constant energy shift, calculated from the training set, is applied to their predicted total energy. Our testing results are summarized in Table.~\ref{tab:mda_dens}.

\begin{table}[htb]
\centering
\caption{Comparison of different methods in terms of the prediction errors for energy, force, dipole, and density, for the malonaldehyde dataset. 
Errors are measured in mH for energy, mH/\AA\ for force, Debye for dipole, and $e$ for density. 
For ML-based methods (sGDML, DeePMD, DeePKS), the types of labels used for training are shown in parentheses. 
The term ``rlxd'' stands for the relaxation procedure after training with density.
}
\begin{tabular}{ll|rrrr}
method &    & $\norm{\Delta E}_1$ & $\norm{\Delta F}_1$ & $\norm{\Delta \mu}_2$ & $\norm{\Delta \rho}_1$ \\
\hline
\hline
sGDML  &(w/$\,E,F$)             & 0.10 & 0.59  & -- & -- \\
DeePMD &(w/$\,E,F$)             & 0.13 & 0.69  & -- & -- \\
\hline
HF     &($E + 805.19$)          & 3.29 & 24.1 & 0.66 & 0.58 \\
PBE    &($E -400.33$)           & 1.35 & 7.53 & 0.17 & 0.35 \\
SCAN0  &($E - 522.05$)          & 1.83 & 10.9 & 0.32 & 0.29 \\
\hline
DeePKS &(w/$\,E$)               & 0.067 & 0.44 & 0.10  & 0.50 \\
DeePKS &(w/$\,E,F$)             & 0.034 & 0.18 & 0.10  & 0.39 \\
DeePKS &(w/$\,E,F,\rho$)        & 0.048 & 0.30 & 0.044 & 0.20 \\
DeePKS &(w/$\,E,F,\rho$; rlxd)  & 0.041 & 0.24 & 0.047 & 0.21 \\
\end{tabular}
\label{tab:mda_dens}
\end{table}

In general, we find ML-based methods perform much better than traditional HF or DFT functionals in terms of the accuracy of energy and forces. 
This is expected since these methods are directly trained with corresponding labels on this specific system. Traditional functionals, on the other hand, give rather good prediction on dipoles and densities. 
Only by including density labels can DeePKS outperform the state-of-the-art conventional functional (SCAN0).
It is also interesting to observe that even without dipole labels, the DeePKS models, obtained in  different ways, significantly outperform HF, PBE, and SCAN0 in terms of testing accuracy on dipole moments.

\begin{figure}[htb]
    \centering
    \includegraphics[width=0.6\textwidth]{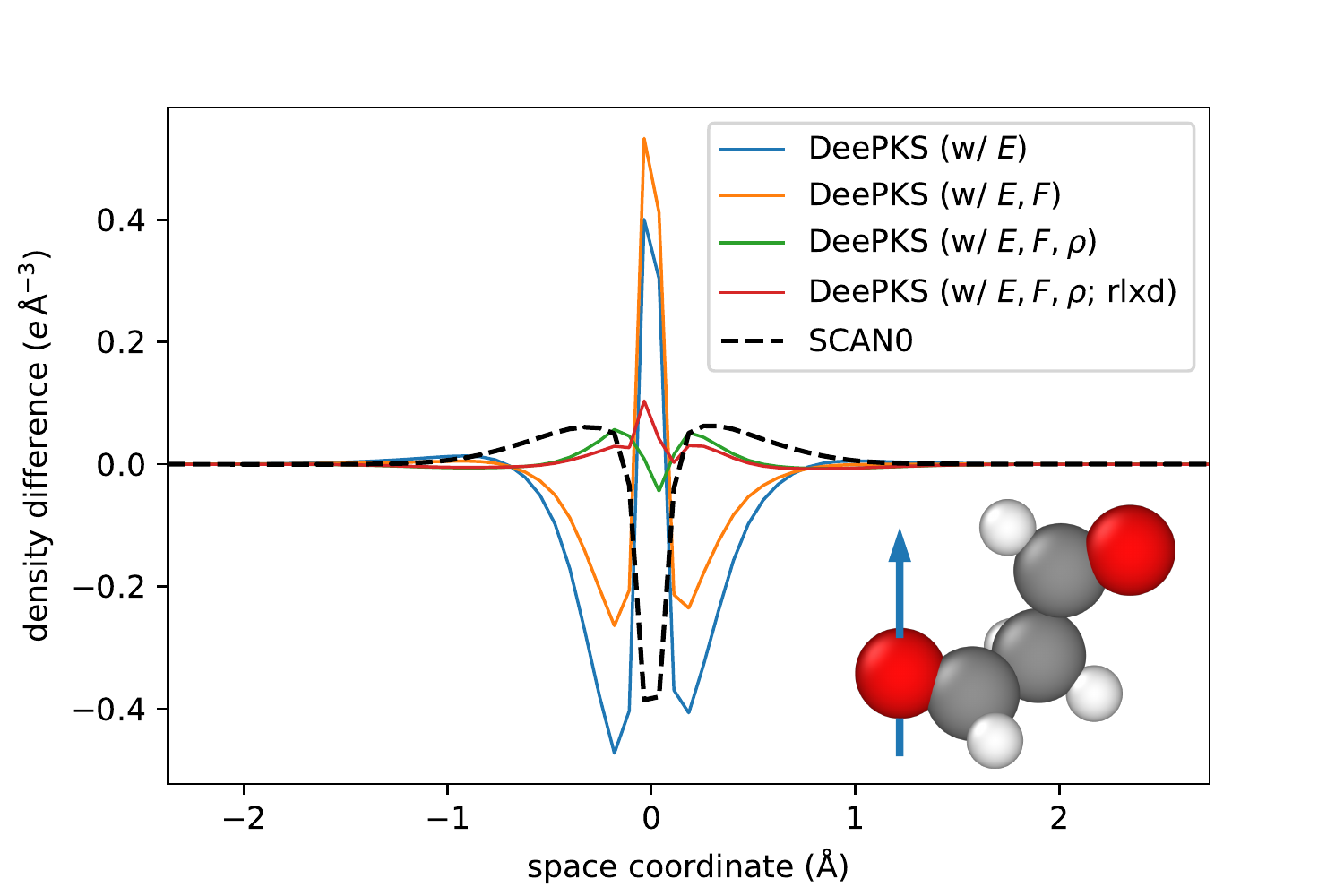}
    \caption{Density difference with respect to $\rho_\mr{CCSD}$ on a sliced line crossing an oxygen atom in the malonaldehyde molecule, given by different training schemes for DeePKS and the SCAN0 functional. The $x$ axis corresponds to space coordinates on the sliced line, which is shown in the inset as a blue arrow. The molecule is drawn by OVITO.\cite{stukowski2009visualization}
    }
    \label{fig:mda_dens}
\end{figure}

For a more intuitive view, we compare the ground-state density given by SCAN0  with different training schemes for DeePKS. 
We take a sliced line that crosses an oxygen atom, and we plot the density difference compared with the CCSD label.
As shown in Fig.~\ref{fig:mda_dens}, when training without density, the error is relatively large (around $0.5\ e\,\mr{\AA}^{-3}$ at maximum) in the core region and is worse than the SCAN0 prediction. 
After we add density labels, the error is  reduced to below $0.1\ e\,\mr{\AA}^{-3}$, showing the necessity of using density labels in the training. 
We also note that the absolute density value can reach $600\ e\,\mr{\AA}^{-3}$ at  the core, hence even the largest difference in density is still very small compare to the absolute value.

As a further step, we test the performance of  DeePKS  on learning one \emph{single} functional for multiple molecules simultaneously. 
This is in general a hard task, especially when the number of training samples is very limited. 
As mentioned in Ref.~\citenum{sauceda2019molecular}, for these so-called transferable models, ``energy prediction errors are often much larger than 1 kcal/mol'', even with huge amount of training data \cite{smith2017ani,smith2018less,smith2019approaching}. However,  this step crucial and inevitable since our ultimate goal is to build \emph{one} universally accurate functional for a wide range of systems.

\begin{table}[htb]
    \centering
    \caption{Comparison of different methods on the prediction accuracy of energy and force for the dataset containing configurations of malonaldehyde, benzene and toluene.
    Errors are measured in mH for energy and mH/\AA\ for force. 
    Force errors are calculated component-wisely. 
    Methods marked with ``*'' are trained separately on each molecules. 
    NeuralXC results are digitally captured from Ref.~\citenum{dick2020machine}.}
    \begin{tabular}{l|rr|rr|rr}
         & \multicolumn{2}{c|}{Malonaldehyde} & \multicolumn{2}{c}{Benzene} & \multicolumn{2}{|c}{Toluene} \\
        method &  $\norm{\Delta E}_1$ & $\norm{\Delta F}_1$ & $\norm{\Delta E}_1$ & $\norm{\Delta F}_1$ & $\norm{\Delta E}_1$ & $\norm{\Delta F}_1$ \\
        \hline
        \hline
        NeuralXC* & 0.35  &   -- & 0.075  &  -- & 0.20  &   -- \\
        sGDML*    & 0.10  & 0.59 & 0.006 & 0.06 & 0.05  & 0.33 \\
        DeePKS*    & 0.04  & 0.22 & 0.007 & 0.07 & 0.06  & 0.32 \\
        DeePKS    & 0.07  & 0.41 & 0.014 & 0.13 & 0.08  & 0.42 \\
    \end{tabular}
    \label{tab:3mol}
\end{table}

We then check the behavior of DeePKS for fitting malonaldehyde, benzene and toluene \emph{at the same time}, with energy and force labels. 
This is the largest set of data we find with both energy and force at the CCSD(T) level calculated in the same numerical setup. 
We take 1000 samples for each molecule in the training and test on the remaining configurations. 
We summarize our results in Table~\ref{tab:3mol}, including a comparison with NeuralXC and sGDML. 
Despite a small loss in accuracy, DeePKS method is still comparable with sGDML and outperforms NeuralXC, both of which are trained separately on each individual molecule. 
We also note that sGDML performs relatively well on benzene and toluene, possibly due to their explicit handling of the point group symmetry. 
Such treatment can improve the sample efficiency  for highly symmetric molecules like benzene and toluene, yet may not be very helpful for more general molecules. 

\begin{figure}[htb]
    \centering
    \includegraphics[width=0.6\textwidth]{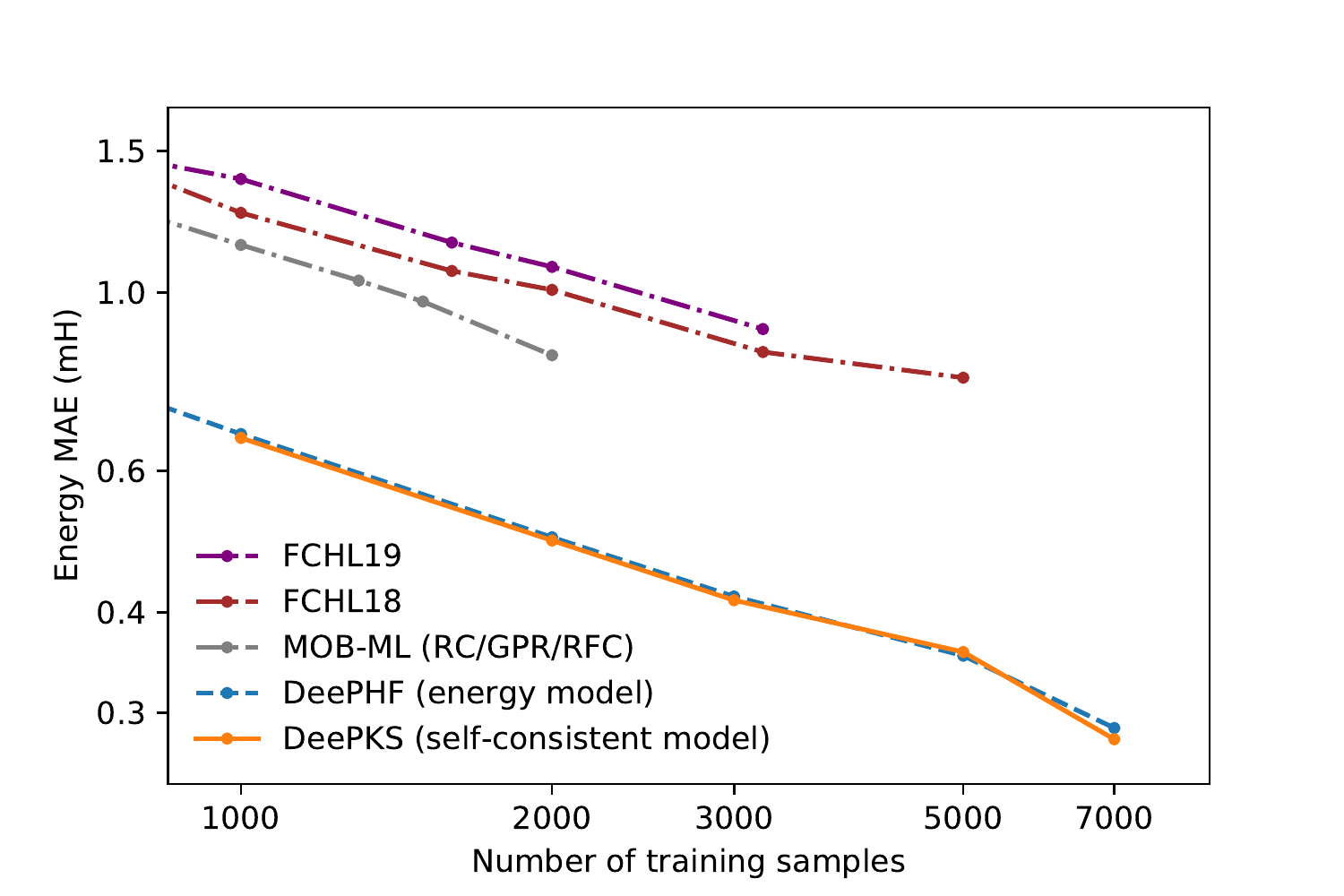}
    \caption{The learning curve of DeePHF and DeePKS methods on the QM7b-T dataset. 
    Results of MOB-ML with regression clustering \cite{cheng2019regression} and FCHL \cite{christensen2020fchl} methods are included for comparison. 
    The FCHL results use MP2 energy as training and testing labels, and are digitally captured from Ref.~\citenum{christensen2020fchl}. }
    \label{fig:qm7_lc}
\end{figure}

For a larger test, we examine the performance of DeePKS on the QM7b-T dataset. 
This is the largest dataset we have with CCSD(T) level of energy, and is also used for benchmarking the energy model  DeePHF \cite{chen2020ground}. 
We study the learning curve by randomly selecting some samples as training set and test on the rest. 
Since there is no new label included and the model is trained only with energy, we should not expect DeePKS to exhibit any accuracy improvement with respect to DeePHF. 
The best results we can look for is that the self-consistent model behaves as well as the energy model. This is indeed the case, as shown in Fig.~\ref{fig:qm7_lc}. 

\begin{table}[htb]
\centering
\caption{MAE of reaction and isomerization energies, calculated using the HC7\cite{peverati2011generalized} and ISOL6\cite{luo2011validation} datasets, respectively.
Results of ANI-1ccx are taken from Ref.~\citenum{smith2019approaching}. 
Errors are given in mH.
All DFT methods' results are obtained using the cc-pVDZ basis. 
The MAEs of DFT methods (including DeePKS) are calculated by comparing with results from CCSD(T)/cc-pVDZ calculation.
The MAE of ANI-1ccx is calculated by comparing with the methods used for generating their training data, i.e., CCSD(T)*/CBS.
}
\begin{tabular}{l|rr}
Methods        & HC7    & ISOL6  \\
\hline
\hline
PBE            & 8.91   & 3.80   \\
SCAN           & 16.22  & 3.25   \\
B3LYP          & 16.74  & 4.16   \\
SCAN0          & 23.94  & 3.65   \\
$\omega$B97X   & 17.71  & 3.45   \\
$\omega$B97M-V & 5.86   & 3.81   \\
\hline
ANI-1ccx       & 3.24   & 2.41   \\
DeePKS         & 2.88   & 1.26   \\
\end{tabular}
\label{tab:ani_abs}
\end{table}

We further examine the transferability of DeePKS to much larger systems by predicting hydrocarbon reaction and isomerization energies using the HC7\cite{peverati2011generalized} and ISOL6\cite{luo2011validation} benchmarks.
The 7000 samples randomly selected from the QM7b-T dataset, used to train the DeePKS model, contain at most 7 heavy atoms.
However, HC7 and ISOL6 contain at most 12 and 15 heavy atoms, respectively.
As shown in Table~\ref{tab:ani_abs}, DeePKS outperforms conventional DFT functionals and generalizes better than the current best-performing ML-based model, ANI1-ccx, which is trained using a huge dataset of 5M molecular configurations with DFT energies and forces, and fine-tuned on about 500K configurations with CCSD(T)*/CBS energies.

\begin{figure}[htb]
    \centering
    \includegraphics[width=0.6\textwidth]{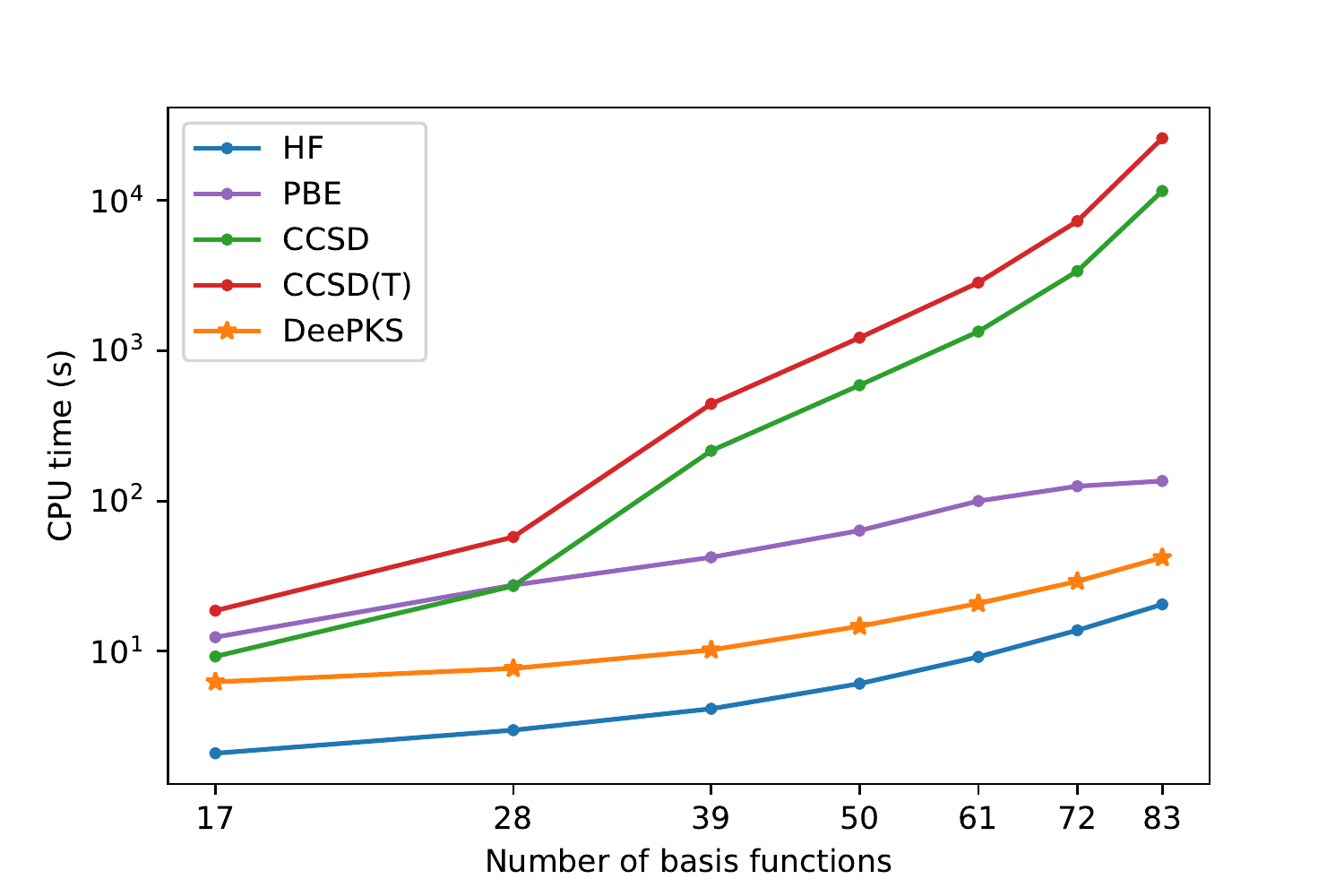}
    \caption{The CPU time spent in the calculations of alkanes using different methods. }
    \label{fig:scaling}
\end{figure}

As a final remark, we show that the DeePKS model can indeed be evaluated efficiently. 
 Fig.~\ref{fig:scaling} shows the computational cost of different methods for calculating alkanes ranging from one to seven carbon atoms.
The number of iterations in all SCF-based methods is set to 10. %
We note that for PBE and other DFT functionals, the implementation in PySCF involves numerical integration over space grids, which is much more expensive for small molecules with the GTO basis set, wherein analytical evaluations of orbital overlapping can be carried out efficiently in the HF method and 
the HF-based DeePKS.  %
As a result,  DeePKS is even faster than PBE and scales similarly with HF.  The additional cost  over HF scales essentially linearly with respect to system size. For larger systems where the $\order{N^4}$ scaling in HF begins to dominate, we can switch to PBE or other KS functionals as the starting point and implement our method in a planewave framework, to retain the cubic scaling. The planewave implementation of DeePKS is left for future work.

\section{Conclusion}
We presented a general framework for learning  chemically accurate self-consistent energy functionals using different types of labels, including energy, force, and density. 
The new training method, combined with a self-consistent extension of DeePHF, leads  to a generalized Kohn-Sham functional with the accuracy of CCSD(T) and the computational cost of DFT. 
We examined the performance of the proposed method on multiple molecular datasets, and 
obtained highly accurate predictions for multiple properties like energy, force, and density.  
In addition, the proposed method is capable of learning a single functional that covers different molecular systems, and its accuracy can be continuously improved by adding more training data. 
We believe it is a good starting point towards a universally accurate functional for molecules, and we are confident that it
can be extended to include condensed phases.

\section{Acknowledgement}
We thank Xiao Wang and Lin Lin for beneficial discussions. 
The work of Y. C., L. Z. and W. E was supported in part by a gift from iFlytek to Princeton University, the ONR grant N00014-13-1-0338, and the Center Chemistry in Solution and at Interfaces (CSI) funded by the DOE Award DE-SC0019394.
The work of H. W. is supported by the National Science Foundation of China under Grant No. 11871110, the National Key Research and Development Program of China under Grants No. 2016YFB0201200 and No. 2016YFB0201203, and Beijing Academy of Artificial Intelligence (BAAI).

\appendix 
\section{Properties of the modified minimization scheme for density optimization}

We discuss the properties of the energy and density when we modify in Eq.~\ref{eqn:emin_dens} the minimization scheme for density optimization.
For simplicity, let us use the following notation:
\begin{equation}
\label{eqn:e-lambda}
 L^{\lambda_\rho}\qty\big[\{\varphi_i\}]=E_\mr{model} \qty\big[\{\varphi_i\}] + \lambda_\rho D\qty\big[\rho[\{\varphi_i\}], \rho_\mr{label}] ,
\end{equation}
and
\begin{equation}
\label{eqn:phi-lambda}
 \{\varphi_i^{\lambda_\rho}\}=\argmin_{ \braket{\varphi_i}{\varphi_j} = \delta_{ij}} L^{\lambda_\rho}\qty\big[\{\varphi_i\}],
\end{equation}
for which we assume that the global minimizer of $L^{\lambda_\rho}\qty\big[\{\varphi_i\}]$ is unique. In particular, 
\begin{equation}
\label{eqn:phi-star}
 \{\varphi_i^*\}=\argmin_{ \braket{\varphi_i}{\varphi_j} = \delta_{ij}} L^0\qty\big[\{\varphi_i\}]
                =\argmin_{ \braket{\varphi_i}{\varphi_j} = \delta_{ij}} E_\mr{model}\qty\big[\{\varphi_i\}],
\end{equation}
gives the original minimizer.

For $\lambda_1>\lambda_2\geq 0$, we have 
\begin{align}
\label{eqn:e-lambda-ineq1}
 L^{\lambda_1}\qty\big[\{\varphi_i^{\lambda_2}\}]&\geq
 L^{\lambda_1}\qty\big[\{\varphi_i^{\lambda_1}\}];\\
\label{eqn:e-lambda-ineq2}
 L^{\lambda_1}\qty\big[\{\varphi_i^{\lambda_1}\}]&\geq
 L^{\lambda_2}\qty\big[\{\varphi_i^{\lambda_1}\}];\\
 \label{eqn:e-lambda-ineq3}
 L^{\lambda_2}\qty\big[\{\varphi_i^{\lambda_1}\}]&\geq 
 L^{\lambda_2}\qty\big[\{\varphi_i^{\lambda_2}\}].
\end{align}

Eq.~\ref{eqn:e-lambda-ineq1} holds, since $\{\varphi_i^{\lambda_1}\}$ is the minimizer of $L^{\lambda_1}$;
 Similarly, Eq.~\ref{eqn:e-lambda-ineq3} holds, since $\{\varphi_i^{\lambda_2}\}$ is the minimizer of $L^{\lambda_2}$;
 Eq.~\ref{eqn:e-lambda-ineq2} holds, since the term $(\lambda_1-\lambda_2) D\qty\big[\rho[\{\varphi_i^{\lambda_1}\}], \rho_\mr{label}]$ is non-negative.

It is straightforward to see that equalities hold for all these equations if and only if both $D\qty\big[\rho[\{\varphi_i^{\lambda_1}\}], \rho_\mr{label}]$ 
and
$D\qty\big[\rho[\{\varphi_i^{\lambda_2}\}], \rho_\mr{label}]$  are 0.
In this case, both the energy $L^{\lambda_\rho}\qty\big[\{\varphi_i^{\lambda_\rho}\}]$ and the minimizing density $\rho[\{\varphi_i^{\lambda_\rho}\}]$ will be the same for all $\lambda_{\rho}\geq0$.
Otherwise, we will have the following two properties:
\begin{enumerate}
    \item $E_\mr{model}\qty\big[\{\varphi_i^{\lambda_\rho}\}]$ is strictly larger than $E_\mr{model}\qty\big[\{\varphi_i^*\}]$, by taking $\lambda_1 = \lambda_\rho$ and $\lambda_2 = 0$ in Eq.~\ref{eqn:e-lambda-ineq3}. 
    \item A larger penalty will lead to a density that is closer to the label. This can be obtained by adding Eq.~\ref{eqn:e-lambda-ineq1} to Eq.~\ref{eqn:e-lambda-ineq3}, which will lead to 
    \begin{equation}
\label{eqn:e-lambda-ineq4}
    (\lambda_1-\lambda_2) (D\qty\big[\rho[\{\varphi_i^{\lambda_2}\}], \rho_\mr{label}]-D\qty\big[\rho[\{\varphi_i^{\lambda_1}\}], \rho_\mr{label}]) \geq 0.
    \end{equation}
    Therefore, we have
    $D\bqty\big{\rho[\{\varphi_i^{\lambda_1}\}], \rho_\mr{label}} \leq D\bqty\big{\rho[\{\varphi_i^{\lambda_2}\}], \rho_\mr{label}}$. 
\end{enumerate}

\section{Visualization of molecular orbitals of malonaldehyde}

\begin{figure}[htbp]
    \centering
    \begin{subfigure}{0.35\textwidth}
    \includegraphics[width=\textwidth]{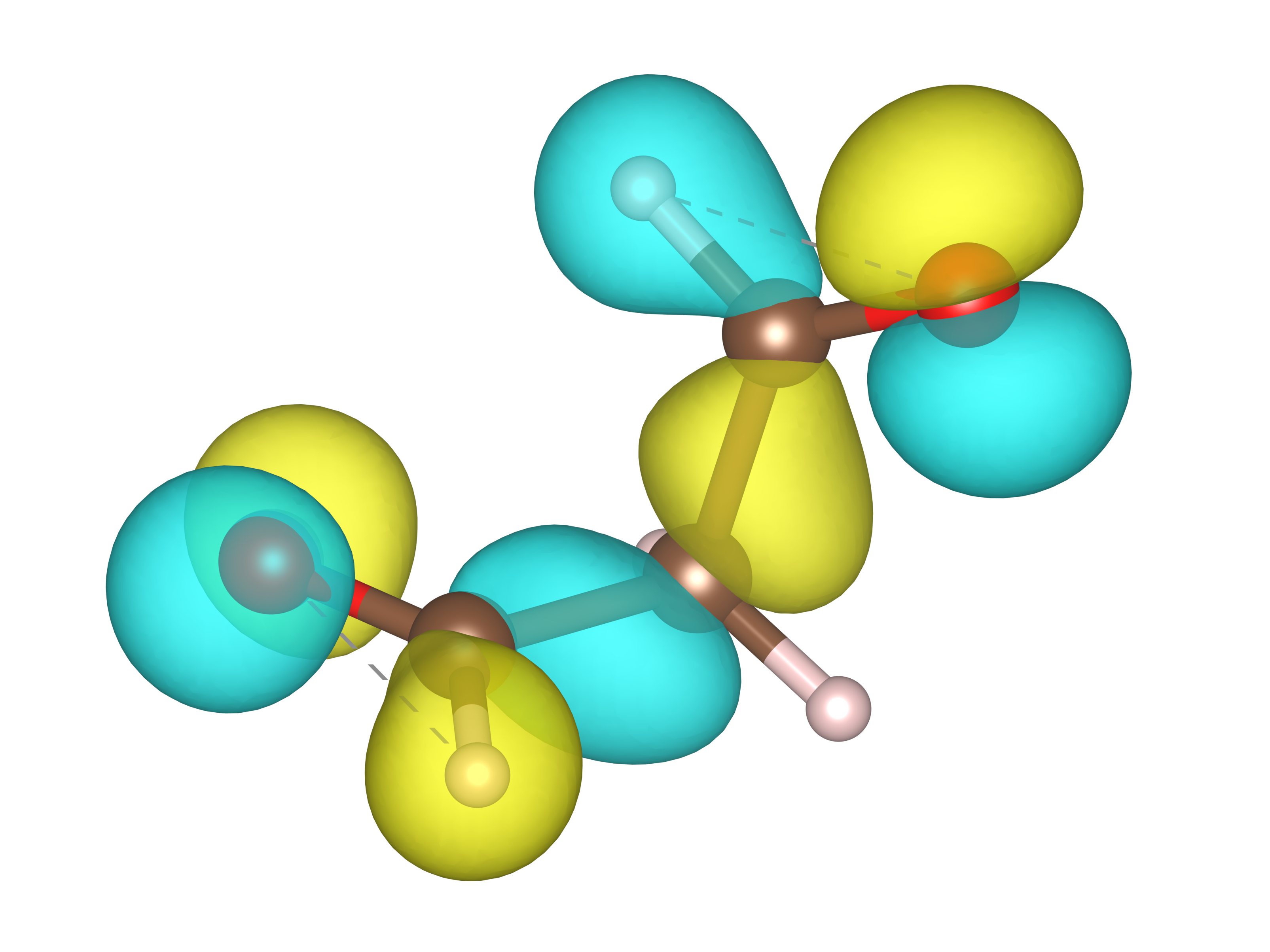}
    \subcaption{HF}
    \end{subfigure}
    \begin{subfigure}{0.35\textwidth}
    \includegraphics[width=\textwidth]{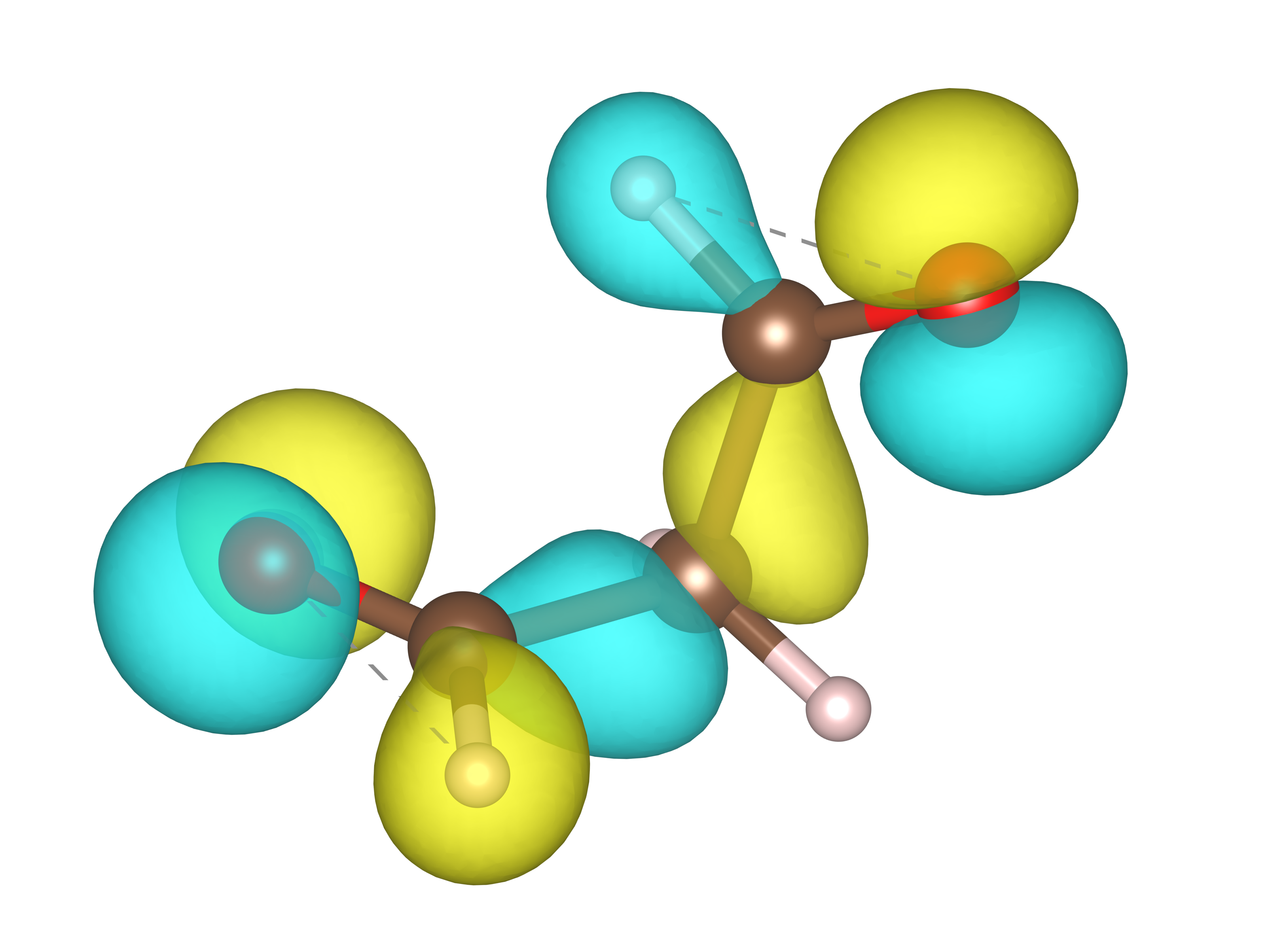}
    \subcaption{PBE}
    \end{subfigure}
    \begin{subfigure}{0.35\textwidth}
    \includegraphics[width=\textwidth]{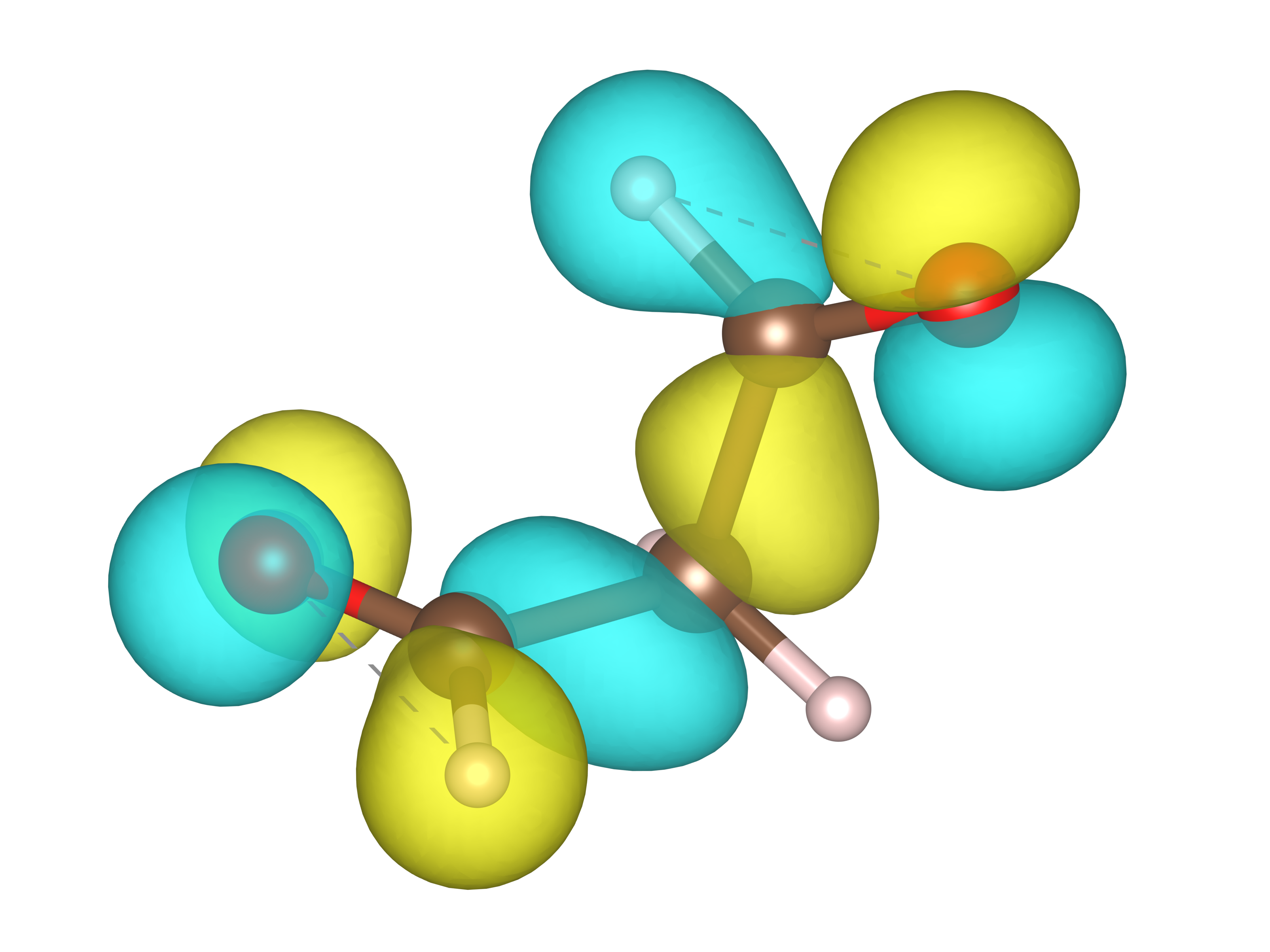}
    \subcaption{DeePKS based on HF}
    \end{subfigure}
    \begin{subfigure}{0.35\textwidth}
    \includegraphics[width=\textwidth]{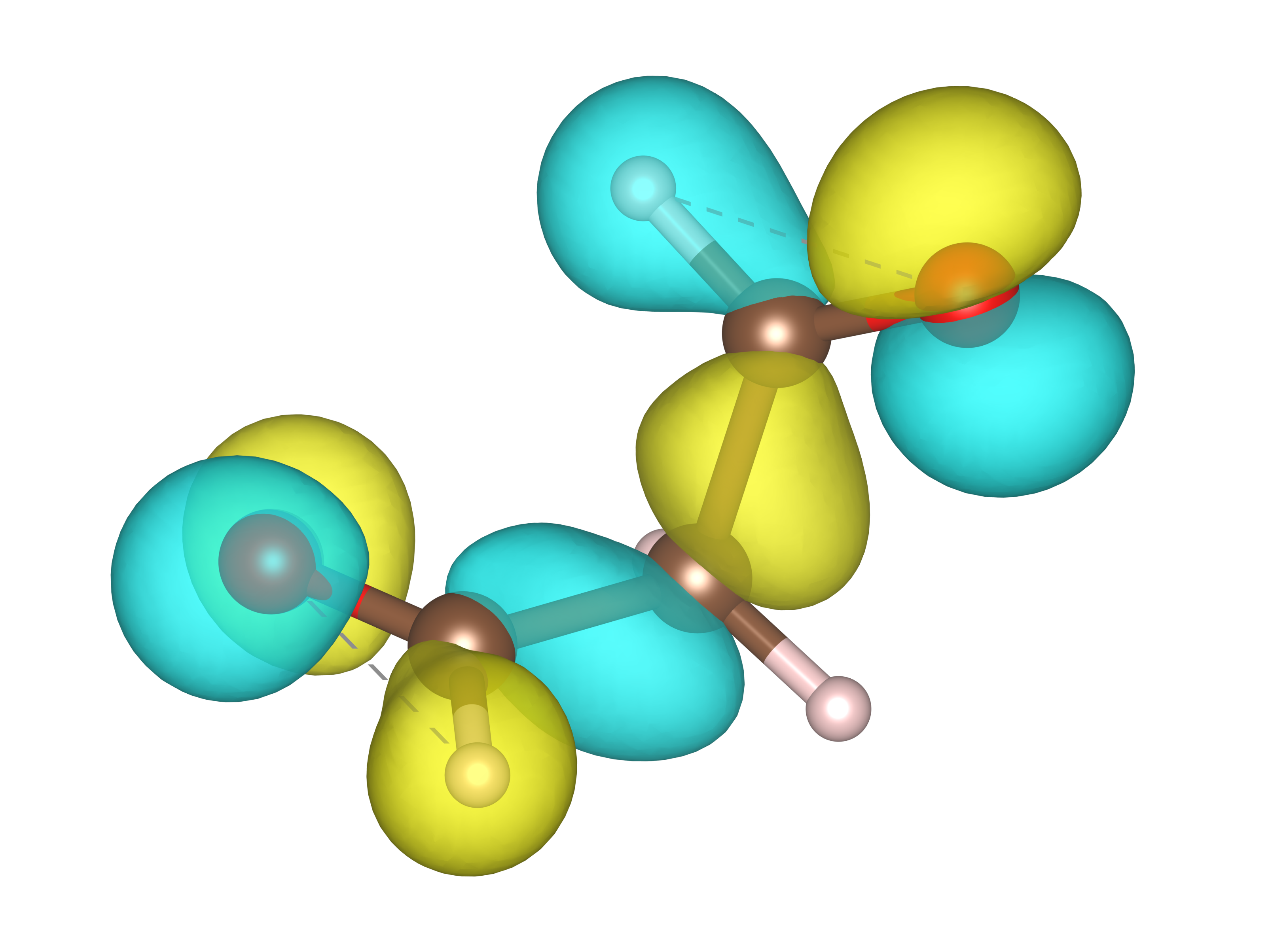}
    \subcaption{DeePKS based on PBE}
    \end{subfigure}
    \caption{HOMO of a typical configuration of malonaldehyde molecule calculated by four different models. The isosurfaces are drawn at the level of 0.05 a.u.. Visualization is done by the VESTA software\cite{momma2011vesta}.}
    \label{fig:molorbs}
\end{figure}

For an intuitive picture on how DeePKS  works, we provide in Fig.~\ref{fig:molorbs} a comparison plot for the highest occupied molecular orbital (HOMO) of malonaldehyde molecule, calculated by four different models, including HF theory, KSDFT with PBE functional, DeePKS based on HF and DeePKS  based on PBE. It can be seen that the two DeePKS models behave similarly. The difference between the two DeePKS models is much smaller than that between the methods they base on, i.e. HF and PBE. This is well expected, since DeePKS approximates the ``exact'' functional that gives the same prediction of its labeling method (CCSD in this case), and should be insensitive to its starting point. We note again  that the orbitals predicted by DeePKS models have no physical meaning. They are shown here as an indication of the robustness of the DeePKS method.

\section{Integrated absolute density difference of malonaldehyde.}

\begin{figure}[htbp]
    \centering
    \begin{subfigure}{0.45\textwidth}
    \includegraphics[width=\textwidth]{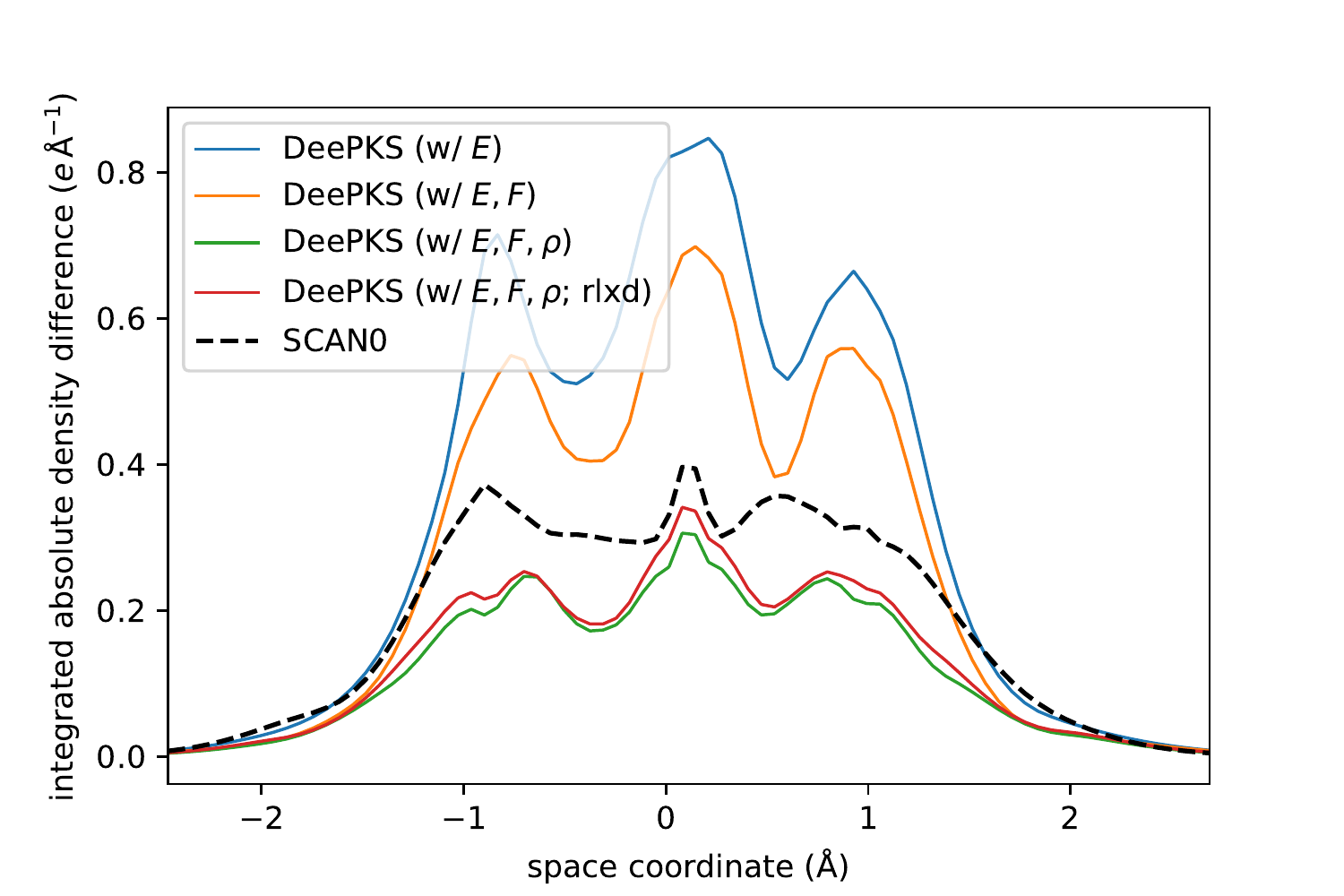}
    \end{subfigure}
    \begin{subfigure}{0.45\textwidth}
    \includegraphics[width=\textwidth]{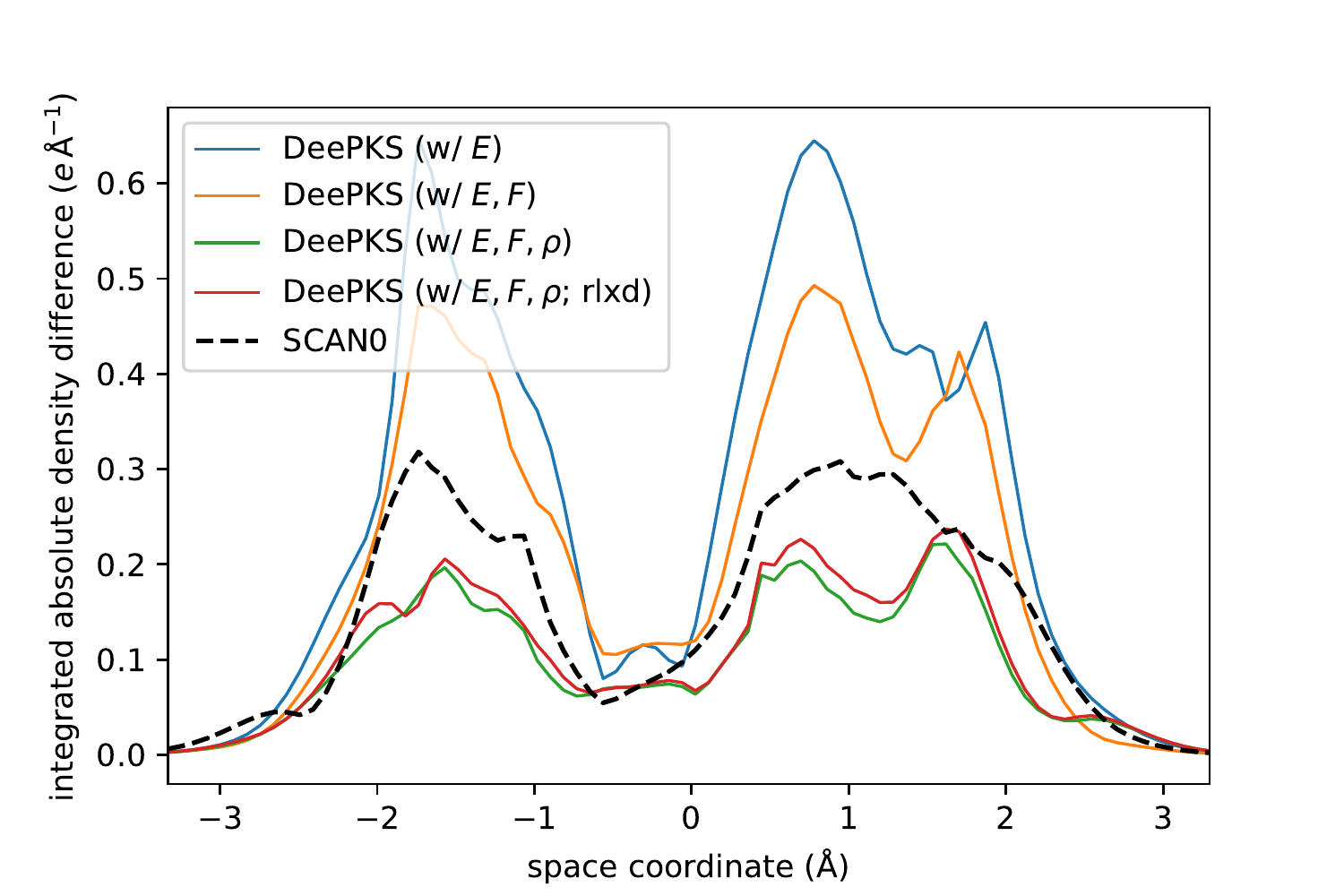}
    \end{subfigure}
    \begin{subfigure}{0.45\textwidth}
    \includegraphics[width=\textwidth]{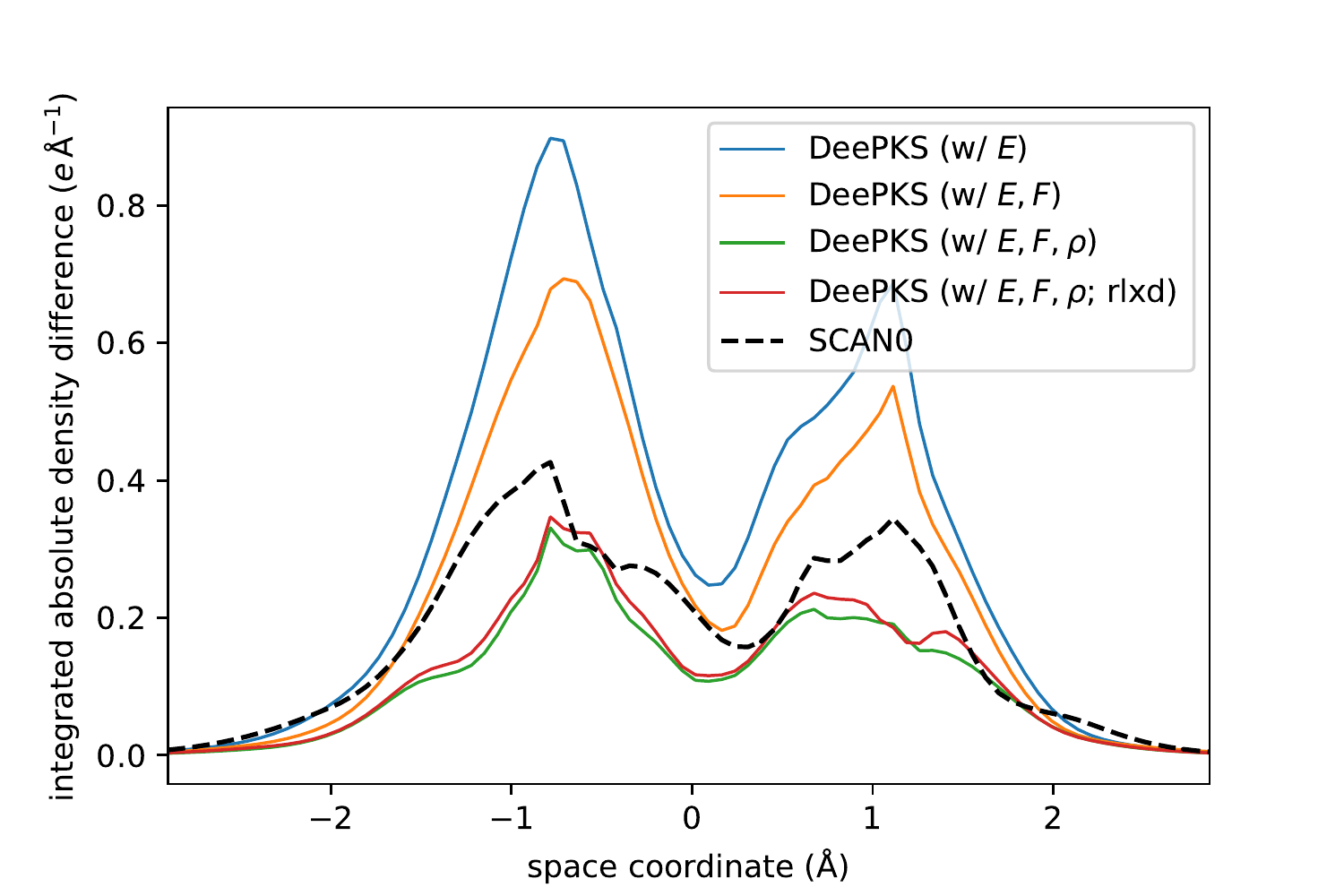}
    \end{subfigure}
    \caption{Integrated absolute density difference of a typical configuration of the malonaldehyde molecule. The absolute density difference is integrated on two spacial dimensions and plotted against the third dimension. The sub-figures corresponds to different integrated space dimensions. }
    \label{fig:mda_intdens}
\end{figure}

We show in Fig.~\ref{fig:mda_intdens} the integrated absolute difference of density calculated by different models. The difference is plotted against one spacial direction with the other two integrated. The findings are similar to the ones shown in Fig.~\ref{fig:mda_dens}. In all cases, the error in density from DeePKS models can be largely reduced by using density as labels in the training procedure. The models trained with density labels can give more accurate density prediction than the SCAN0 functional.

\section{Learning curve of three molecules}

\begin{figure}[htbp]
    \centering
    \includegraphics[width=0.6\textwidth]{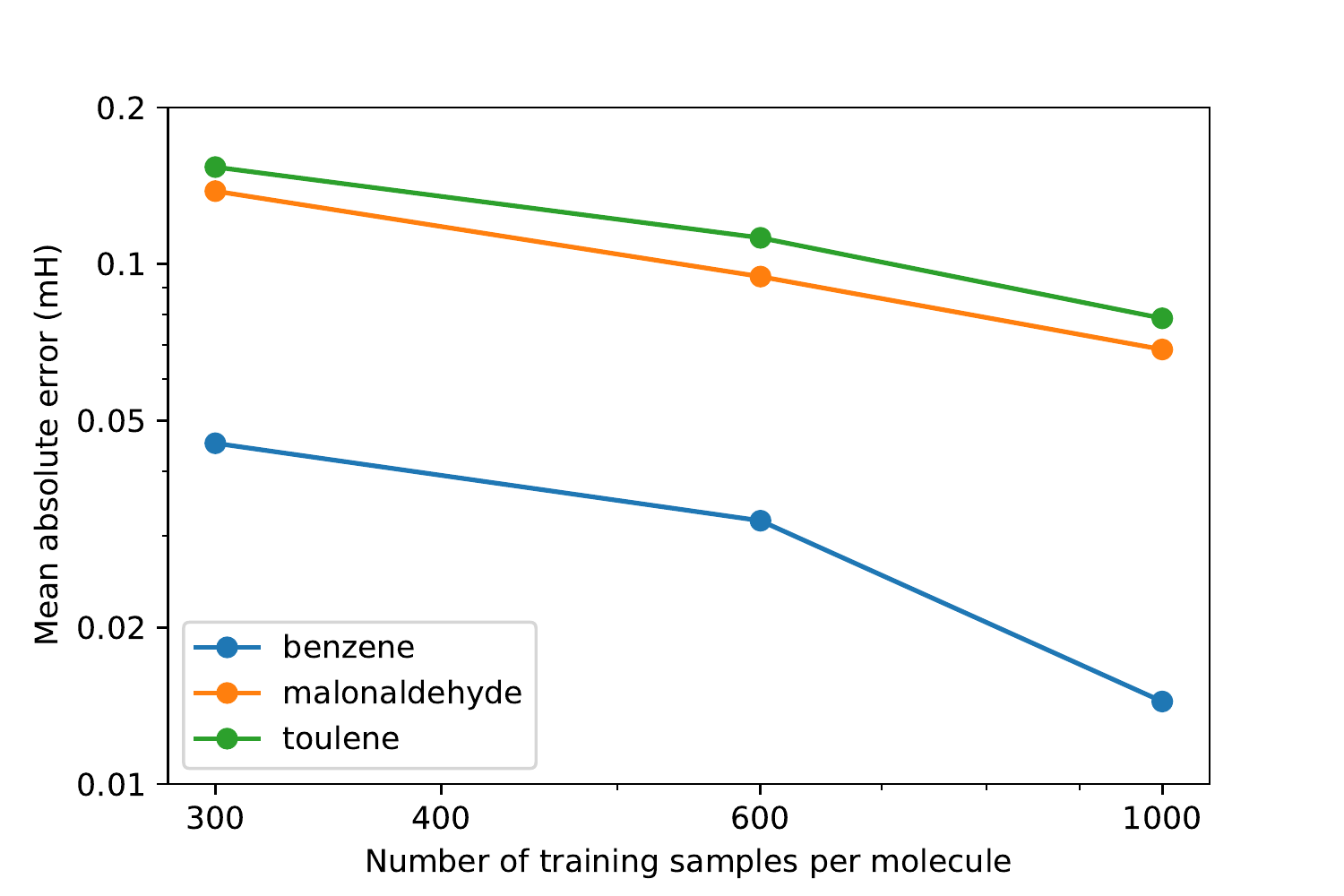}
    \caption{The learning curve of the DeePKS method on the dataset containing snapshots of malonaldehyde, benzene and toulene molecules. 
    For each choice of the training data size, a single DeePKS model is trained for all  three molecules simultaneously.}
    \label{fig:m3_lc}
\end{figure}

We provide in Fig.~\ref{fig:m3_lc} the learning curve of DeePKS trained on the dataset containing snapshots of malonaldehyde, benzene and toulene molecules at the same time. 

\bibliographystyle{unsrtnat}  %
\bibliography{ref_abbr}

\begin{thebibliography}{50}
\providecommand{\natexlab}[1]{#1}
\providecommand{\url}[1]{\texttt{#1}}
\expandafter\ifx\csname urlstyle\endcsname\relax
  \providecommand{\doi}[1]{doi: #1}\else
  \providecommand{\doi}{doi: \begingroup \urlstyle{rm}\Url}\fi

\bibitem[Hohenberg and Kohn(1964)]{hohenberg1964inhomogeneous}
Pierre Hohenberg and Walter Kohn.
\newblock Inhomogeneous electron gas.
\newblock \emph{Phys. Rev.}, 136\penalty0 (3B):\penalty0 B864, 1964.

\bibitem[Pople et~al.(1987)Pople, Head-Gordon, and Raghavachari]{pople1987QCI}
John~A Pople, Martin Head-Gordon, and Krishnan Raghavachari.
\newblock Quadratic configuration interaction. {A} general technique for
  determining electron correlation energies.
\newblock \emph{J. Chem. Phys.}, 87\penalty0 (10):\penalty0 5968--5975, 1987.

\bibitem[Jeziorski and Monkhorst(1981)]{jeziorski1981coupled}
Bogumil Jeziorski and Hendrik~J Monkhorst.
\newblock Coupled-cluster method for multideterminantal reference states.
\newblock \emph{Phys. Rev. A}, 24\penalty0 (4):\penalty0 1668, 1981.

\bibitem[Kohn and Sham(1965)]{kohn1965self}
Walter Kohn and Lu~Jeu Sham.
\newblock Self-consistent equations including exchange and correlation effects.
\newblock \emph{Phys. Rev.}, 140\penalty0 (4A):\penalty0 A1133, 1965.

\bibitem[Seidl et~al.(1996)Seidl, G{\"o}rling, Vogl, Majewski, and
  Levy]{seidl1996generalized}
A~Seidl, Andreas G{\"o}rling, Peter Vogl, Jacek~A Majewski, and Mel Levy.
\newblock Generalized kohn-sham schemes and the band-gap problem.
\newblock \emph{Phys. Rev. B}, 53\penalty0 (7):\penalty0 3764, 1996.

\bibitem[Perdew et~al.(1996)Perdew, Burke, and
  Ernzerhof]{perdew1996generalized}
John~P Perdew, Kieron Burke, and Matthias Ernzerhof.
\newblock Generalized gradient approximation made simple.
\newblock \emph{Phys. Rev. Lett.}, 77\penalty0 (18):\penalty0 3865, 1996.

\bibitem[Sun et~al.(2015)Sun, Ruzsinszky, and Perdew]{sun2015SCAN}
Jianwei Sun, Adrienn Ruzsinszky, and John~P Perdew.
\newblock Strongly constrained and appropriately normed semilocal density
  functional.
\newblock \emph{Phys. Rev. Lett.}, 115\penalty0 (3):\penalty0 036402, 2015.

\bibitem[Chen et~al.(2020)Chen, Zhang, Wang, and E]{chen2020ground}
Yixiao Chen, Linfeng Zhang, Han Wang, and Weinan E.
\newblock Ground state energy functional with hartree--fock efficiency and
  chemical accuracy.
\newblock \emph{J. Phys. Chem. A}, 124\penalty0 (35):\penalty0 7155--7165,
  2020.

\bibitem[Snyder et~al.(2012)Snyder, Rupp, Hansen, M{\"u}ller, and
  Burke]{snyder2012finding}
John~C Snyder, Matthias Rupp, Katja Hansen, Klaus-Robert M{\"u}ller, and Kieron
  Burke.
\newblock Finding density functionals with machine learning.
\newblock \emph{Phys. Rev. Lett.}, 108\penalty0 (25):\penalty0 253002, 2012.

\bibitem[Bogojeski et~al.(2019)Bogojeski, Vogt-Maranto, Tuckerman, Mueller, and
  Burke]{bogojeski2019density}
Mihail Bogojeski, Leslie Vogt-Maranto, Mark~E Tuckerman, Klaus-Robert Mueller,
  and Kieron Burke.
\newblock Density functionals with quantum chemical accuracy: From machine
  learning to molecular dynamics.
\newblock \emph{ChemRxiv preprint}, 8079917:\penalty0 v1, 2019.

\bibitem[Dick and Fernandez-Serra(2020)]{dick2020machine}
Sebastian Dick and Marivi Fernandez-Serra.
\newblock Machine learning accurate exchange and correlation functionals of the
  electronic density.
\newblock \emph{Nat. Commun.}, 11\penalty0 (1):\penalty0 1--10, 2020.

\bibitem[Lei and Medford(2019)]{lei2019design}
Xiangyun Lei and Andrew~J Medford.
\newblock Design and analysis of machine learning exchange-correlation
  functionals via rotationally invariant convolutional descriptors.
\newblock \emph{Phys. Rev. Mater.}, 3\penalty0 (6):\penalty0 063801, 2019.

\bibitem[Liu et~al.(2017)Liu, Wang, Du, Hu, Zheng, and Chen]{liu2017imporving}
Qin Liu, JingChun Wang, PengLi Du, LiHong Hu, Xiao Zheng, and GuanHua Chen.
\newblock Improving the performance of long-range-corrected
  exchange-correlation functional with an embedded neural network.
\newblock \emph{J. Phys. Chem. A}, 121\penalty0 (38):\penalty0 7273--7281,
  2017.

\bibitem[Nagai et~al.(2020)Nagai, Akashi, and Sugino]{nagai2020completing}
Ryo Nagai, Ryosuke Akashi, and Osamu Sugino.
\newblock Completing density functional theory by machine learning hidden
  messages from molecules.
\newblock \emph{npj Comput. Mater.}, 6\penalty0 (1):\penalty0 1--8, 2020.

\bibitem[Levy(1979)]{levy1979universal}
Mel Levy.
\newblock Universal variational functionals of electron densities, first-order
  density matrices, and natural spin-orbitals and solution of the
  v-representability problem.
\newblock \emph{Proc. Natl. Acad. Sci. U. S. A.}, 76\penalty0 (12):\penalty0
  6062--6065, 1979.

\bibitem[Lieb(1983)]{lieb1983density}
Elliott~H Lieb.
\newblock Density functionals for coulomb systems.
\newblock \emph{Int. J. Quantum Chem.}, 24\penalty0 (3):\penalty0 243--277,
  1983.

\bibitem[Gilbert(1975)]{gilbert1975hohenberg}
Thomas~L Gilbert.
\newblock Hohenberg-kohn theorem for nonlocal external potentials.
\newblock \emph{Phys. Rev. B}, 12\penalty0 (6):\penalty0 2111, 1975.

\bibitem[M{\o}ller and Plesset(1934)]{moller1934note}
Chr M{\o}ller and Milton~S Plesset.
\newblock Note on an approximation treatment for many-electron systems.
\newblock \emph{Phys. Rev.}, 46\penalty0 (7):\penalty0 618, 1934.

\bibitem[Cheng et~al.(2019{\natexlab{a}})Cheng, Welborn, Christensen, and
  Miller~III]{cheng2019universal}
Lixue Cheng, Matthew Welborn, Anders~S Christensen, and Thomas~F Miller~III.
\newblock A universal density matrix functional from molecular orbital-based
  machine learning: Transferability across organic molecules.
\newblock \emph{J. Chem. Phys.}, 150\penalty0 (13):\penalty0 131103,
  2019{\natexlab{a}}.

\bibitem[Han et~al.(2019)Han, Zhang, and Weinan]{han2019solving}
Jiequn Han, Linfeng Zhang, and E~Weinan.
\newblock Solving many-electron schr{\"o}dinger equation using deep neural
  networks.
\newblock \emph{J. Comput. Phys.}, 399:\penalty0 108929, 2019.

\bibitem[Hermann et~al.(2020)Hermann, Sch{\"a}tzle, and
  No{\'e}]{hermann2020deep}
Jan Hermann, Zeno Sch{\"a}tzle, and Frank No{\'e}.
\newblock Deep-neural-network solution of the electronic schr{\"o}dinger
  equation.
\newblock \emph{Nat. Chem.}, pages 1--7, 2020.

\bibitem[Pfau et~al.(2020)Pfau, Spencer, Matthews, and Foulkes]{pfau2020ab}
David Pfau, James~S Spencer, Alexander~GDG Matthews, and W~Matthew~C Foulkes.
\newblock Ab initio solution of the many-electron schr{\"o}dinger equation with
  deep neural networks.
\newblock \emph{Phys. Rev. Res.}, 2\penalty0 (3):\penalty0 033429, 2020.

\bibitem[Behler and Parrinello(2007)]{behler2007generalized}
J{\"o}rg Behler and Michele Parrinello.
\newblock Generalized neural-network representation of high-dimensional
  potential-energy surfaces.
\newblock \emph{Phys. Rev. Lett.}, 98\penalty0 (14):\penalty0 146401, 2007.

\bibitem[Bart{\'o}k et~al.(2010)Bart{\'o}k, Payne, Kondor, and
  Cs{\'a}nyi]{bartok2010gaussian}
Albert~P Bart{\'o}k, Mike~C Payne, Risi Kondor, and G{\'a}bor Cs{\'a}nyi.
\newblock Gaussian approximation potentials: The accuracy of quantum mechanics,
  without the electrons.
\newblock \emph{Phys. Rev. Lett.}, 104\penalty0 (13):\penalty0 136403, 2010.

\bibitem[Rupp et~al.(2012)Rupp, Tkatchenko, M{\"u}ller, and
  VonLilienfeld]{rupp2012fast}
Matthias Rupp, Alexandre Tkatchenko, Klaus-Robert M{\"u}ller, and O~Anatole
  VonLilienfeld.
\newblock Fast and accurate modeling of molecular atomization energies with
  machine learning.
\newblock \emph{Phys. Rev. Lett.}, 108\penalty0 (5):\penalty0 058301, 2012.

\bibitem[Ramakrishnan et~al.(2015)Ramakrishnan, Dral, Rupp, and von
  Lilienfeld]{ramakrishnan2015big}
Raghunathan Ramakrishnan, Pavlo~O Dral, Matthias Rupp, and O~Anatole von
  Lilienfeld.
\newblock Big data meets quantum chemistry approximations: The $\delta$-machine
  learning approach.
\newblock \emph{J. Chem. Theory Comput.}, 11\penalty0 (5):\penalty0 2087--2096,
  2015.

\bibitem[Chmiela et~al.(2017)Chmiela, Tkatchenko, Sauceda, Poltavsky,
  Sch{\"u}tt, and M{\"u}ller]{chmiela2017machine}
Stefan Chmiela, Alexandre Tkatchenko, Huziel~E Sauceda, Igor Poltavsky,
  Kristof~T Sch{\"u}tt, and Klaus-Robert M{\"u}ller.
\newblock Machine learning of accurate energy-conserving molecular force
  fields.
\newblock \emph{Sci. Adv.}, 3\penalty0 (5):\penalty0 e1603015, 2017.

\bibitem[Sch{\"u}tt et~al.(2017)Sch{\"u}tt, Kindermans, Felix, Chmiela,
  Tkatchenko, and M{\"u}ller]{schutt2017schnet}
Kristof Sch{\"u}tt, Pieter-Jan Kindermans, Huziel Enoc~Sauceda Felix, Stefan
  Chmiela, Alexandre Tkatchenko, and Klaus-Robert M{\"u}ller.
\newblock Schnet: A continuous-filter convolutional neural network for modeling
  quantum interactions.
\newblock \emph{Adv. Neural Inf. Process. Syst.}, pages 992--1002, 2017.

\bibitem[Smith et~al.(2017)Smith, Isayev, and Roitberg]{smith2017ani}
Justin~S Smith, Olexandr Isayev, and Adrian~E Roitberg.
\newblock {ANI-1}: an extensible neural network potential with dft accuracy at
  force field computational cost.
\newblock \emph{Chem. Sci.}, 8\penalty0 (4):\penalty0 3192--3203, 2017.

\bibitem[Han et~al.(2018)Han, Zhang, Car, and E]{han2017deep}
Jiequn Han, Linfeng Zhang, Roberto Car, and Weinan E.
\newblock Deep potential: a general representation of a many-body potential
  energy surface.
\newblock \emph{Commun. Comput. Phys.}, 23\penalty0 (3):\penalty0 629--639,
  2018.

\bibitem[Zhang et~al.(2018{\natexlab{a}})Zhang, Han, Wang, Car, and
  E]{zhang2018deep}
Linfeng Zhang, Jiequn Han, Han Wang, Roberto Car, and Weinan E.
\newblock Deep potential molecular dynamics: A scalable model with the accuracy
  of quantum mechanics.
\newblock \emph{Phys. Rev. Lett.}, 120:\penalty0 143001, Apr
  2018{\natexlab{a}}.

\bibitem[Zhang et~al.(2018{\natexlab{b}})Zhang, Han, Wang, Saidi, Car, and
  E]{zhang2018end}
Linfeng Zhang, Jiequn Han, Han Wang, Wissam Saidi, Roberto Car, and Weinan E.
\newblock End-to-end symmetry preserving inter-atomic potential energy model
  for finite and extended systems.
\newblock \emph{Adv. Neural Inf. Process. Syst.}, pages 4436--4446,
  2018{\natexlab{b}}.

\bibitem[Brockherde et~al.(2017)Brockherde, Vogt, Li, Tuckerman, Burke, and
  M{\"u}ller]{brockherde2017bypassing}
Felix Brockherde, Leslie Vogt, Li~Li, Mark~E Tuckerman, Kieron Burke, and
  Klaus-Robert M{\"u}ller.
\newblock Bypassing the kohn-sham equations with machine learning.
\newblock \emph{Nat. Commun.}, 8\penalty0 (1):\penalty0 1--10, 2017.

\bibitem[Grisafi et~al.(2018)Grisafi, Fabrizio, Meyer, Wilkins, Corminboeuf,
  and Ceriotti]{grisafi2018transferable}
Andrea Grisafi, Alberto Fabrizio, Benjamin Meyer, David~M Wilkins, Clemence
  Corminboeuf, and Michele Ceriotti.
\newblock Transferable machine-learning model of the electron density.
\newblock \emph{ACS Cent. Sci.}, 5\penalty0 (1):\penalty0 57--64, 2018.

\bibitem[Chandrasekaran et~al.(2019)Chandrasekaran, Kamal, Batra, Kim, Chen,
  and Ramprasad]{chandrasekaran2019solving}
Anand Chandrasekaran, Deepak Kamal, Rohit Batra, Chiho Kim, Lihua Chen, and
  Rampi Ramprasad.
\newblock Solving the electronic structure problem with machine learning.
\newblock \emph{npj Comput. Mater.}, 5\penalty0 (1):\penalty0 1--7, 2019.

\bibitem[Zepeda-N{\'u}{\~n}ez et~al.(2019)Zepeda-N{\'u}{\~n}ez, Chen, Zhang,
  Jia, Zhang, and Lin]{zepeda2019deep}
Leonardo Zepeda-N{\'u}{\~n}ez, Yixiao Chen, Jiefu Zhang, Weile Jia, Linfeng
  Zhang, and Lin Lin.
\newblock Deep density: circumventing the kohn-sham equations via symmetry
  preserving neural networks.
\newblock \emph{arXiv preprint}, page 1912.00775, 2019.

\bibitem[Friedman(2001)]{friedman2001greedy}
Jerome~H Friedman.
\newblock Greedy function approximation: a gradient boosting machine.
\newblock \emph{Ann. Math. Stat.}, pages 1189--1232, 2001.

\bibitem[Sun et~al.(2018)Sun, Berkelbach, Blunt, Booth, Guo, Li, Liu, McClain,
  Sayfutyarova, Sharma, et~al.]{sun2018pyscf}
Qiming Sun, Timothy~C Berkelbach, Nick~S Blunt, George~H Booth, Sheng Guo,
  Zhendong Li, Junzi Liu, James~D McClain, Elvira~R Sayfutyarova, Sandeep
  Sharma, et~al.
\newblock Pyscf: the python-based simulations of chemistry framework.
\newblock \emph{Wiley Interdiscip. Rev.: Comput. Mol. Sci.}, 8\penalty0
  (1):\penalty0 e1340, 2018.

\bibitem[Sauceda et~al.(2019)Sauceda, Chmiela, Poltavsky, M{\"u}ller, and
  Tkatchenko]{sauceda2019molecular}
Huziel~E Sauceda, Stefan Chmiela, Igor Poltavsky, Klaus-Robert M{\"u}ller, and
  Alexandre Tkatchenko.
\newblock Molecular force fields with gradient-domain machine learning:
  Construction and application to dynamics of small molecules with coupled
  cluster forces.
\newblock \emph{J. Chem. Phys.}, 150\penalty0 (11):\penalty0 114102, 2019.

\bibitem[Cheng et~al.()Cheng, Welborn, Christensen, and
  Miller~III]{cheng2019data}
Lixue Cheng, Matthew Welborn, Anders~S Christensen, and Thomas~F Miller~III.
\newblock Thermalized (350k) qm7b, gdb-13, water, and short alkane quantum
  chemistry dataset including mob-ml features.
\newblock \url{https://data.caltech.edu/records/1177} (accessed July 7, 2020).

\bibitem[Cheng et~al.(2019{\natexlab{b}})Cheng, Kovachki, Welborn, and
  Miller~III]{cheng2019regression}
Lixue Cheng, Nikola~B Kovachki, Matthew Welborn, and Thomas~F Miller~III.
\newblock Regression clustering for improved accuracy and training costs with
  molecular-orbital-based machine learning.
\newblock \emph{J. Chem. Theory Comput.}, 15\penalty0 (12):\penalty0
  6668--6677, 2019{\natexlab{b}}.

\bibitem[Christensen et~al.(2020)Christensen, Bratholm, Faber, and Anatole~von
  Lilienfeld]{christensen2020fchl}
Anders~S Christensen, Lars~A Bratholm, Felix~A Faber, and O~Anatole~von
  Lilienfeld.
\newblock Fchl revisited: Faster and more accurate quantum machine learning.
\newblock \emph{J. Chem. Phys.}, 152\penalty0 (4):\penalty0 044107, 2020.

\bibitem[Paszke et~al.(2019)Paszke, Gross, Massa, Lerer, Bradbury, Chanan,
  Killeen, Lin, Gimelshein, Antiga, Desmaison, Kopf, Yang, DeVito, Raison,
  Tejani, Chilamkurthy, Steiner, Fang, Bai, and Chintala]{PyTorch2019}
Adam Paszke, Sam Gross, Francisco Massa, Adam Lerer, James Bradbury, Gregory
  Chanan, Trevor Killeen, Zeming Lin, Natalia Gimelshein, Luca Antiga, Alban
  Desmaison, Andreas Kopf, Edward Yang, Zachary DeVito, Martin Raison, Alykhan
  Tejani, Sasank Chilamkurthy, Benoit Steiner, Lu~Fang, Junjie Bai, and Soumith
  Chintala.
\newblock Pytorch: An imperative style, high-performance deep learning library.
\newblock \emph{Adv. Neural Inf. Process. Syst.}, pages 8024--8035, 2019.

\bibitem[Kingma and Ba(2014)]{kingma2014adam}
Diederik~P Kingma and Jimmy Ba.
\newblock Adam: A method for stochastic optimization.
\newblock \emph{arXiv preprint}, page 1412.6980, 2014.

\bibitem[Stukowski(2009)]{stukowski2009visualization}
Alexander Stukowski.
\newblock Visualization and analysis of atomistic simulation data with
  ovito--the open visualization tool.
\newblock \emph{Modell. Simul. Mater. Sci. Eng.}, 18\penalty0 (1):\penalty0
  015012, 2009.

\bibitem[Smith et~al.(2018)Smith, Nebgen, Lubbers, Isayev, and
  Roitberg]{smith2018less}
Justin~S Smith, Ben Nebgen, Nicholas Lubbers, Olexandr Isayev, and Adrian~E
  Roitberg.
\newblock Less is more: Sampling chemical space with active learning.
\newblock \emph{J. Chem. Phys.}, 148\penalty0 (24):\penalty0 241733, 2018.

\bibitem[Smith et~al.(2019)Smith, Nebgen, Zubatyuk, Lubbers, Devereux, Barros,
  Tretiak, Isayev, and Roitberg]{smith2019approaching}
Justin~S Smith, Benjamin~T Nebgen, Roman Zubatyuk, Nicholas Lubbers, Christian
  Devereux, Kipton Barros, Sergei Tretiak, Olexandr Isayev, and Adrian~E
  Roitberg.
\newblock Approaching coupled cluster accuracy with a general-purpose neural
  network potential through transfer learning.
\newblock \emph{Nat. Commun.}, 10\penalty0 (1):\penalty0 1--8, 2019.

\bibitem[Peverati et~al.(2011)Peverati, Zhao, and
  Truhlar]{peverati2011generalized}
Roberto Peverati, Yan Zhao, and Donald~G Truhlar.
\newblock Generalized gradient approximation that recovers the second-order
  density-gradient expansion with optimized across-the-board performance.
\newblock \emph{J. Phys. Chem. Lett.}, 2\penalty0 (16):\penalty0 1991--1997,
  2011.

\bibitem[Luo et~al.(2011)Luo, Zhao, and Truhlar]{luo2011validation}
Sijie Luo, Yan Zhao, and Donald~G Truhlar.
\newblock Validation of electronic structure methods for isomerization
  reactions of large organic molecules.
\newblock \emph{Phys. Chem. Chem. Phys.}, 13\penalty0 (30):\penalty0
  13683--13689, 2011.

\bibitem[Momma and Izumi(2011)]{momma2011vesta}
Koichi Momma and Fujio Izumi.
\newblock Vesta 3 for three-dimensional visualization of crystal, volumetric
  and morphology data.
\newblock \emph{J. Appl. Crystallogr.}, 44\penalty0 (6):\penalty0 1272--1276,
  2011.

\end{thebibliography}

\end{document}